\documentclass[twoside,twocolumn,9pt]{article}
\usepackage{extsizes}
\usepackage{epstopdf}
\usepackage[super,sort&compress,comma]{natbib} 
\usepackage[version=3]{mhchem}
\usepackage[left=1.5cm, right=1.5cm, top=1.785cm, bottom=2.0cm]{geometry}
\usepackage{balance}
\usepackage{times,mathptmx}
\usepackage{sectsty}
\usepackage{graphicx} 
\usepackage{lastpage}
\usepackage[format=plain,justification=justified,singlelinecheck=false,font={stretch=1.125,small,sf},labelfont=bf,labelsep=space]{caption}
\usepackage{float}
\usepackage{fancyhdr}
\usepackage{fnpos}
\usepackage[english]{babel}
\addto{\captionsenglish}{%
  
}
\usepackage{array}
\usepackage{droidsans}
\usepackage{charter}
\usepackage[T1]{fontenc}
\usepackage[usenames,dvipsnames]{xcolor}
\usepackage{setspace}
\usepackage[compact]{titlesec}
\usepackage{hyperref}

\usepackage{epstopdf}

\definecolor{cream}{RGB}{222,217,201}

\begin{document}

\pagestyle{fancy}
\thispagestyle{plain}
\fancypagestyle{plain}{

\fancyhead[C]{\includegraphics[width=18.5cm]{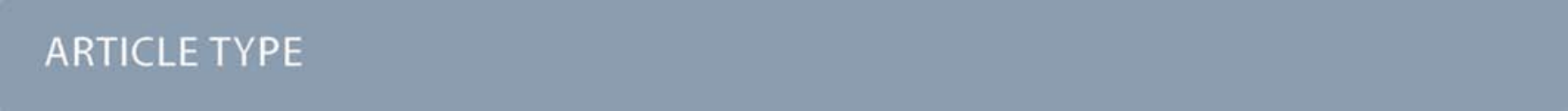}}
\fancyhead[L]{\hspace{0cm}\vspace{1.5cm}\includegraphics[height=30pt]{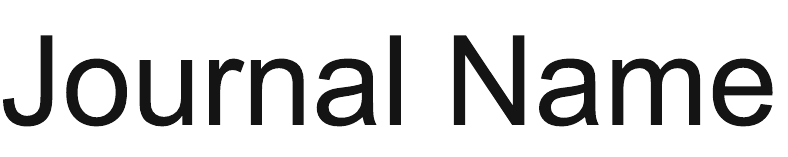}}
\fancyhead[R]{\hspace{0cm}\vspace{1.7cm}\includegraphics[height=55pt]{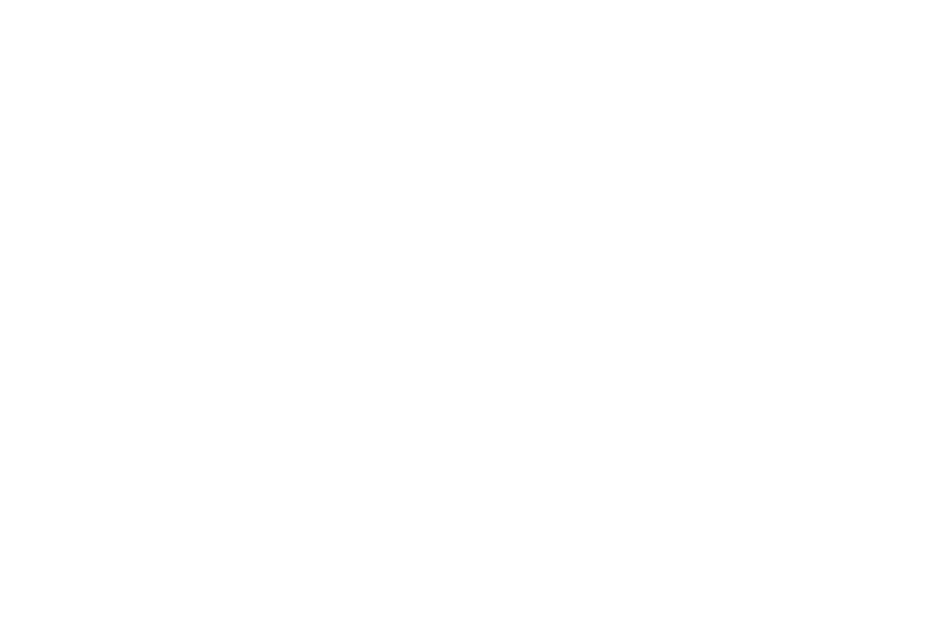}}
\renewcommand{\headrulewidth}{0pt}
}

\makeFNbottom
\makeatletter
\renewcommand\LARGE{\@setfontsize\LARGE{15pt}{17}}
\renewcommand\Large{\@setfontsize\Large{12pt}{14}}
\renewcommand\large{\@setfontsize\large{10pt}{12}}
\renewcommand\footnotesize{\@setfontsize\footnotesize{7pt}{10}}
\makeatother

\renewcommand{\thefootnote}{\fnsymbol{footnote}}
\renewcommand\footnoterule{\vspace*{1pt}%
\color{cream}\hrule width 3.5in height 0.4pt \color{black}\vspace*{5pt}} 
\setcounter{secnumdepth}{5}

\makeatletter 
\renewcommand\@biblabel[1]{#1}            
\renewcommand\@makefntext[1]%
{\noindent\makebox[0pt][r]{\@thefnmark\,}#1}
\makeatother 
\renewcommand{\figurename}{\small{Fig.}~}
\sectionfont{\sffamily\Large}
\subsectionfont{\normalsize}
\subsubsectionfont{\bf}
\setstretch{1.125} 
\setlength{\skip\footins}{0.8cm}
\setlength{\footnotesep}{0.25cm}
\setlength{\jot}{10pt}
\titlespacing*{\section}{0pt}{4pt}{4pt}
\titlespacing*{\subsection}{0pt}{15pt}{1pt}

\fancyfoot{}
\fancyfoot[LO,RE]{\vspace{-7.1pt}\includegraphics[height=9pt]{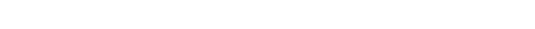}}
\fancyfoot[CO]{\vspace{-7.1pt}\hspace{13.2cm}\includegraphics{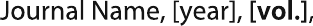}}
\fancyfoot[CE]{\vspace{-7.2pt}\hspace{-14.2cm}\includegraphics{head_foot/RF}}
\fancyfoot[RO]{\footnotesize{\sffamily{1--\pageref{LastPage} ~\textbar  \hspace{2pt}\thepage}}}
\fancyfoot[LE]{\footnotesize{\sffamily{\thepage~\textbar\hspace{3.45cm} 1--\pageref{LastPage}}}}
\fancyhead{}
\renewcommand{\headrulewidth}{0pt} 
\renewcommand{\footrulewidth}{0pt}
\setlength{\arrayrulewidth}{1pt}
\setlength{\columnsep}{6.5mm}
\setlength\bibsep{1pt}

\makeatletter 
\newlength{\figrulesep} 
\setlength{\figrulesep}{0.5\textfloatsep} 

\newcommand{\topfigrule}{\vspace*{-1pt}%
\noindent{\color{cream}\rule[-\figrulesep]{\columnwidth}{1.5pt}} }

\newcommand{\botfigrule}{\vspace*{-2pt}%
\noindent{\color{cream}\rule[\figrulesep]{\columnwidth}{1.5pt}} }

\newcommand{\dblfigrule}{\vspace*{-1pt}%
\noindent{\color{cream}\rule[-\figrulesep]{\textwidth}{1.5pt}} }

\makeatother

\twocolumn[
  \begin{@twocolumnfalse}
\vspace{3cm}
\sffamily
\begin{tabular}{m{4.5cm} p{13.5cm} }

\includegraphics{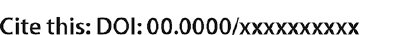} & \noindent\LARGE{\textbf{Long-wavelength fluctuations and static correlations in quasi-2D colloidal suspensions$^\dag$}} \\
\vspace{0.3cm} & \vspace{0.3cm} \\

 & \noindent\large{Bo Zhang,\textit{$^{a}$} and Xiang Cheng \textit{$^{a}$}$^{\ast}$} \\

\includegraphics{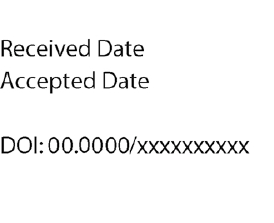} & \noindent\normalsize{Dimensionality strongly affects thermal fluctuations and critical dynamics of equilibrium systems. These influences persist in amorphous systems going through the nonequilibrium glass transition. Here, we experimentally study the glass transition of quasi-2D suspensions of spherical and ellipsoidal particles under different degrees of circular confinement. We show that the strength of the long-wavelength fluctuations increases logarithmically with system sizes and displays the signature of the Mermin-Wagner fluctuations. Moreover, using confinement as a tool, we also measure static structural correlations and extract a growing static correlation length in 2D supercooled liquids. Finally, we explore the influence of the Mermin-Wagner fluctuations on the translational and orientational relaxations of 2D ellipsoidal suspensions, which leads to a new interpretation of the two-step glass transition and the orientational glass phase of anisotropic particles. Our study reveals the importance of long-wavelength fluctuations in 2D supercooled liquids and provides new insights into the role of dimensionality in the glass transition.} \\

\end{tabular}

 \end{@twocolumnfalse} \vspace{0.6cm}

  ]

\renewcommand*\rmdefault{bch}\normalfont\upshape
\rmfamily
\section*{}
\vspace{-1cm}


\footnotetext{\textit{$^{a}$~Department of Chemical Engineering and Materials Science, University of Minnesota, Minneapolis, MN 55455, USA.}}
\footnotetext{$^{\ast}$ Email: xcheng@umn.edu}

\footnotetext{\dag~Electronic Supplementary Information (ESI) available: A document describing detailed data analysis methods is included. See DOI: 00.0000/00000000.}



\section{Introduction}

Thermal fluctuations become increasingly important as the dimensionality of systems is reduced, which fundamentally changes the nature of structural phase transitions in equilibrium systems. In one dimension (1D), thermal fluctuations block the connectivity of systems and therefore destroy any possible long-range orders and finite-temperature phase transitions.\cite{Peierls34} In 2D, the classic Mermin-Wagner theory demonstrates that long-wavelength thermal fluctuations eliminate spontaneous breaking of continuous symmetries with short-range interactions.\cite{Mermin66} Although finite-temperature phase transitions still exist, the nature of the transitions is completely different from their counterparts in 3D. While the long-range translational and rotational orders emerge simultaneously in the discontinuous transition of 3D crystallization, the Kosterlitz, Thouless, Halperin, Nelson, Young (KTHNY) theory shows that the 2D crystallization/melting occurs in two steps continuously via an intermediate hexatic phase, which shows long-range rotational order but no long-range translational order.\cite{Kosterlitz72,Halperin78,Young79}  Glass transition is  intrinsically a nonequilibrium process and infamous for lacking any obvious symmetry breaking. Nevertheless, thermal fluctuations and dimensionality may still strongly affect the transition dynamics. Flenner and Szamel have found in simulations that the relaxation of the translation of particles decouples from the local rearrangements of particles with respect to their neighbors in 2D supercooled liquids, fundamentally different from the dynamics of 3D glass-forming systems.\cite{Flenner15} Later experiments and simulations showed that such a decoupling arises from long-wavelength fluctuations in 2D.\cite{Shiba16,Vivek17,Illing17,Shiba18} Simulations further illustrated that these fluctuations exhibit the same feature as the Mermin-Wagner (MW) fluctuations.\cite{Shiba16,Illing17} However, the nature of these fluctuations has not been tested in experiments. Furthermore, although a two-step glass transition associated with the translational and orientational relaxations of supercooled liquids of anisotropic particles has been identified in 2D,\cite{Zheng11,Zheng14} whether and how these different degrees of freedom of 2D anisotropic supercooled liquids are affected by the observed long-wavelength fluctuations are still open questions.    

In addition to the numerical and experimental findings, thermodynamic theories of the glass transition also predict a possible dependence of structural correlations of supercooled liquids on the dimensionality of systems.\cite{Berthier11,Cammarota12} The classic Adam-Gibbs theory proposes that the super-Arrhenius growth of material relaxation time near the glass transition temperature $T_g$ arises from increasingly large regions of cooperative molecular rearrangement.\cite{Gibbs65,Debenedetti01} The length scale associated with such regions is quantitatively analyzed by the Random First Order Transition (RFOT) theory.\cite{Kirkpatrick89,Biroli08,Cammarota12,Cammarota13,Chakrabarty16} In RFOT, confinement is employed ingeniously as a tool to probe the structural correlations of bulk supercooled liquids.\cite{Biroli08,Cammarota13} An imaginary cavity of size $R$ is created in a bulk supercooled liquid. Particles inside the cavity are allowed to freely evolve to explore different amorphous states, whereas particles outside the cavity are frozen in an equilibrium configuration. The structural correlations between the particles near the center of the cavity and those at the boundary of the cavity can then be quantitatively measured. Numerical simulations revealed a non-zero correlation when $R$ is smaller than a certain length $\xi$. This so-called point-to-set correlation length $\xi$ increases with decreasing temperature $T$.\cite{Biroli08} Later experiments on colloidal suspensions near flat boundaries \cite{Nagamanasa15} and inside cavities of different sizes \cite{Bo16} qualitatively confirmed the numerical finding. More quantitatively, the free-energy cost due to the mismatch between the amorphous state of particles within a cavity of $R$ and the equilibrium state outside $R$ is $\Delta F_{s} = \Gamma R^{\theta}$, where $\Gamma$ is a generalized surface tension with $\theta \leq d-1$ and $d$ is the dimensionality. The free-energy gain from the multiplicity of different amorphous states is $\Delta F_{b} = -Ts_c(T)R^d$, where $s_c$ is the configurational entropy density. $s_c$ vanishes linearly as $T$ approaching $T_K$, $s_c=k(T-T_K)$, where $T_K < T_g$ is the Kauzmann temperature and $k$ is a dimensional constant. Balancing the total free energy leads to a prediction of $\xi$, 
\begin{equation}
\label{equ1} {\xi = \left[\frac{\Gamma}{Ts_c(T)}\right]^{1/(d-\theta)}=\left[\frac{\Gamma(T)}{Tk(T-T_K)}\right]^{1/(d-\theta)}}.
\end{equation} 
While $\theta = 2$ has been reported in 3D simulations,\cite{Cammarota09,Brun12,Chakrabarty16} $\theta$ in 2D is still inconclusive and may be model-dependent.\cite{Chakrabarty16} This controversy regarding the dimensionality dependence can potentially be resolved experimentally by probing the structural correlation of supercooled colloidal liquids at different dimensionalities.     

Motivated by the above questions, we experimentally study the glass transition in quasi-2D colloidal suspensions of various aspect ratios under different degress of circular confinement. Confinement is employed in our study as a tool to accomplish two goals. First, it allows us to investigate long-wavelength fluctuations in 2D systems of different sizes. Our experiments show that the strength of the long-wavelength fluctuations increases logarithmaically with the linear size of systems and displays a defining feature of the Mermin-Wagner fluctuations. Second, confinement is also used to probe static structural correlations in 2D supercooled liquids within the framework of RFOT. We identify a structural correlation length near the ideal glass transition, which shows a divergent power-law scaling consistent with the prediction of RFOT. A further quantitative comparison between the scaling relations in 2D and 3D glass-forming systems provides important insights into the glass transition in different dimensionalites.\cite{Bo16} We finally explore the differential influences of the long-wavelength fluctuations on the translational and orientational relaxations of ellipsoidal particles, which leads to a new interpretation on the two-step glass transition.\cite{Zheng11,Zheng14}                    

\section{Experiments}

\begin{figure}
	\begin{center}
		\includegraphics[width=3.55in]{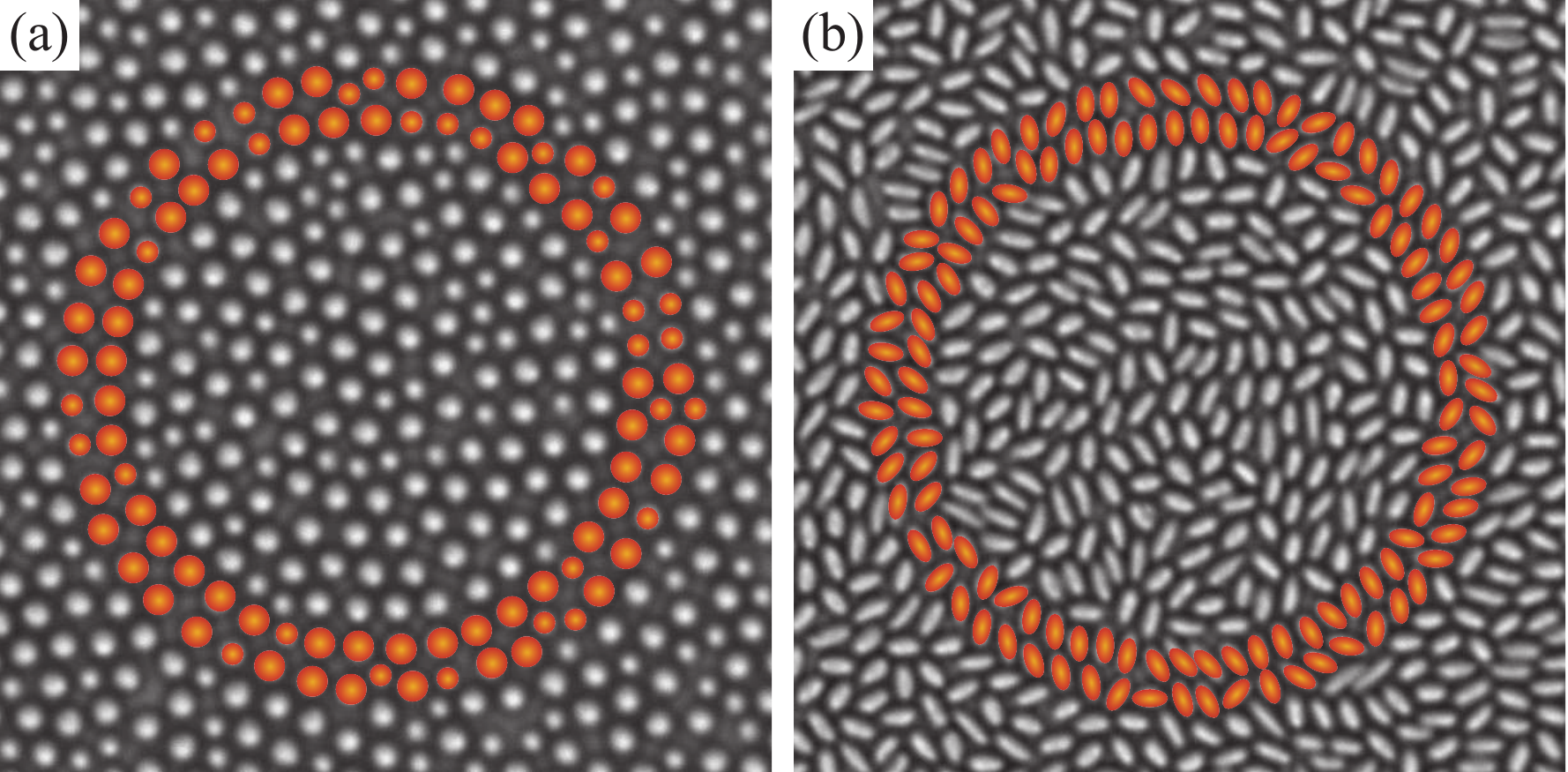}
	\end{center}
	\caption[Microscopic Images of 2D Confined Systems]
	{Confined quasi-2D suspensions of (a) binary spherical particles and (b) ellipsoids. Two layers of particles (marked in orange) are pinned by optical traps. The inner radius of the pinned particle layers is $R = 20$ $\mu$m. The aspect ratio of the ellipsoids is $P=1.9$.  
	}
	\label{Figure1}
\end{figure}

Colloidal particles of different aspect ratios are used to prepare quasi-2D suspensions in our experiments. We first study the structure and dynamics of aqueous quasi-2D colloidal suspensions made of spherical poly(methyl methacrylate) (PMMA) particles of two different sizes. The diameters of the small and the large particles are $d_s = 2.2$ $\mu$m and $d_l = 2.8$ $\mu$m, respectively, which successfully suppress crystallization at high area fractions, $\phi$ (Fig.~\ref{Figure1}a). The polydispersity of each type of particles is less than $5\%$. 7 mM of sodium dodecyl sulfate (SDS) below the critical micelle concentration (CMC) is added to stabilize the suspensions and reduce the Debye length of particles ($<20$ nm).\cite{Zheng11} The suspensions are filled into a wedge-shaped cell. We use 4 $\mu$m silica particles as spacers at the one end of the cell, which lead to a small wedge angle of $8 \times 10^{-5}$. Thus, a large area of quasi-2D colloidal suspensions with almost constant spacing can be selected at a fixed distance away from the end of the cell.\cite{Nugent07} We apply different degrees of circular confinement onto the sample within the quasi-2D plane by pinning a ring of particles at radius $R$ using optical tweezers (Aresis, Tweez 250si) (Fig.~\ref{Figure1}a). The tweezers use a continuous wave infrared laser of 1064 nm with a maximal power of 5 W. The laser power is adjusted accordingly, so that the pinning force on individual particles is independent of $R$. Although two different sizes of particles have been used in our experiments, a certain degree of particle layering can still be observed near the confined wall at high $\phi$, typical for packing of hard spheres near smooth boundaries. \cite{Seidler00} To reduce the influence of these boundary layers, we limit our measurements to the central region of confined samples at least two particle diameters away from the pinned particle layer. Note also that we pin the particles at a fixed distance $R$ in our experiments, different from the disordered equilibrium configurations adopted in RFOT simulations.\cite{Biroli08}   

To investigate the influence of long-wavelength fluctuations on the translational and orientational relaxations of supercooled liquids, we also study the structure and dynamics of quasi-2D suspensions of ellipsoidal particles. Ellipsoids of two different aspect ratios are prepared by mechanical stretching spherical PMMA particles of diameter $d_s = 2.2$ $\mu$m.\cite{Ho93,Yi16} The method for preparing quasi-2D samples of ellipsoidal particles is similar to that for spherical particles. For particles with the small aspect ratio $P = 1.9$, optical tweezers are used to create confined samples of different sizes (Fig.~\ref{Figure1}b). For particles with the large aspect ratio $P = 6.7$, the pinning potential of optical tweezers is flat along the particles' major axis. As a result, high-aspect-ratio particles can move freely inside an optical trap along its major axis due to thermal fluctuations. We are not able to pin the particles using optical tweezers to create confinement. Thus, we focus on the bulk behaviors of the high-aspect-ratio ellipsoids in our experiments. 

\begin{figure}
	\begin{center}
		\includegraphics[width=3.55in]{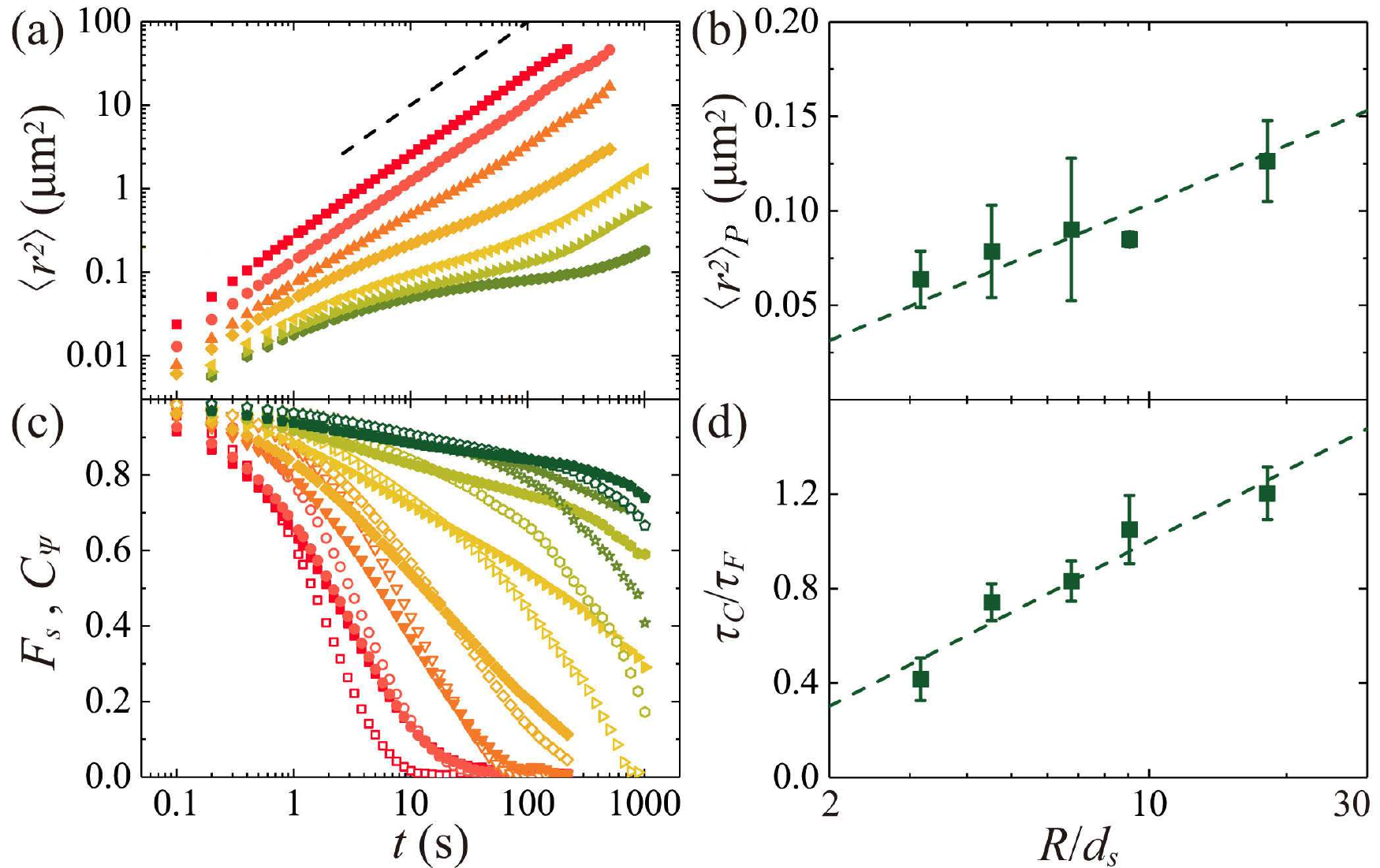}
	\end{center}
	\caption[aa]
	{Long-wavelength fluctuations in quasi-2D supercooled liquids of spherical particles. (a) Mean square displacements of bulk samples, $\langle r^2 \rangle(t)$. From top to bottom, the area fraction $\phi = 0.229, 0.538, 0.695, 0.785, 0.815, 0.822$ and $0.840$. The black dashed line has a slope of 1. (b) Plateaued MSDs at the inflection point, $\langle r^2 \rangle_{P}$, as a function of the system size $R$ for $\phi= 0.82 \pm 0.01$. The dotted line indicates a logarithmic increase. $\langle r^2 \rangle_{P} = 0.15 \pm 0.04$ $\mu$m$^2$ for bulk samples. (c) Correlation functions $F_s(t)$ (open symbols) and $C_\Psi(t)$ (solid symbols) for bulk samples at different $\phi$. From left to right: $\phi = 0.229, 0.534, 0.695, 0.776, 0.814, 0.826, 0.838$ and $0.851$. (d) The relaxation time ratio $\tau_{C}/\tau_{F}$ as a function of $R$ for $\phi = 0.82 \pm 0.01$. The dotted line indicates a logarithmic dependence. $\tau_{C}/\tau_{F} = 9.26 \pm 1.97$ for bulk samples. 
	}
	\label{Figure2}
\end{figure}

\section{Results and discussion}

\subsection{Long-wavelength fluctuations}

We first study the structure and dynamics of spherical PMMA particles in a quasi-2D layer under different degrees of confinement. To reveal the nature of the long-wavelength fluctuations in the 2D supercooled liquids observed in Ref.~7-10, we measure the mean square displacements (MSDs) of particles (Fig.~\ref{Figure2}a). At high $\phi$, MSDs show diffusive behaviors at both short and long times and exhibit the characteristic plateaus at intermediate times, signaling the caging of particles. Since the motion of particles is constrained locally within cages, the plateaued MSDs can be used to quantify the strength of long-wavelength fluctuations at fixed $\phi$.\cite{Shiba16,Illing17} Specifically, we measure plateaued MSDs at the inflection point, $\langle r^2 \rangle_{P}\equiv\langle r^2 \rangle(t=t_f)$, where $(d^2\langle r^2 \rangle/dt^2)|_{t=t_f} = 0$. $\langle r^2 \rangle_{P}$ as a function of the size of confined systems, $R/d_s$, is shown in Fig.~\ref{Figure2}b. At high $\phi = 0.82$, we find that $\langle r^2 \rangle_{P}$ increases logarithmically with $R/d_s$, an unmistaken signature of the Mermin-Wagner (MW) fluctuations.\cite{Mermin66,Dash78,Frohlich81} The magnitude of the change of $\langle r^2 \rangle_{P}$ is consistent with the numerical simulation of the MW fluctuations in 2D colloidal and molecular supercooled liquids.\cite{Shiba16,Illing17} At lower $\phi$, where the distances between nearest neighbors show high variances, $\langle r^2 \rangle_{P}$ shows the characteristic logarithmic increase only under strong confinement but plateaus at large $R/d_s$ (Fig.~\ref{Figure6}a in Appendix A). Although the MW fluctuations occur in high-density amorphous systems in absence of periodicity, they cannot exist in low-density gases.   

The logarithmic dependence and the finite range of confinement limit the dynamic range of $\langle r^2 \rangle_{P}$. Furthermore, the MSD plateaus disappear for sufficiently large samples at a given $\phi$,\cite{Flenner15} preventing the application of $\langle r^2 \rangle_{P}$ in quantifying the long-wavelength fluctuations in very large systems. Hence, we should not rely on $\langle r^2 \rangle_{P}$ as the sole evidence for the MW fluctuations. To further verify the nature of the fluctuations, we quantify their strength via the long-time relaxation of particle dynamics. Particle dynamics with and without the influence of long-wavelength fluctuations can be quantified by the self-intermediate scattering function, $F_s$, and the bond-orientational correlation function, $C_{\Psi}$, respectively:\cite{Vivek17} 
\begin{eqnarray}
F_s(t) & = & \left\langle \frac{1}{N} \sum_{j} \cos\left(\vec{Q} \cdot \lbrack \vec{r}_j(t_0+t) - \vec{r}_j(t_0)\rbrack \right) \right\rangle_{t_0} \\
C_{\Psi}(t) & = & \frac{\left\langle \sum_{j} \lbrack \psi_n^j(t_0)  \rbrack^* \cdot \lbrack \psi_n^j(t_0+t)\rbrack\right\rangle_{t_0}}{\left\langle \sum_j |\psi_n^j(t_0)|^2 \right\rangle_{t_0}}, \label{Cpsi} 
\end{eqnarray}
where $\vec{r}_j(t)$ is the location of particle $j$ at time $t$ and $N$ is the total number of particles. $\vec{Q}=Q\vec{i}_x$ is along the $x$ direction with $Q=5.8/d_s$ obtained from the position of the first peak of the structural factor at high $\phi$. $\psi_n^j(t)= (N_b^j)^{-1}\sum_m\exp(in\theta_{m})$, where $m = 1...N_b^j$ is the nearest neighbor of particle $j$ and $\theta_{m}$ is the angle between the $x$ axis and $\vec{r}_{mj}(t)=\vec{r}_m(t)-\vec{r}_j(t)$. We fix $n=6$ in our study since most particles in our binary samples have six nearest neighbors. Quantitatively similar relaxation is observed when the bond-orientational correlation is calculated by taking average of $n$ from 4 to 8 (Fig.~\ref{Figure7} in Appendix A). Physically, $F_s$ depends on the net displacement of particles and is affected by long-wavelength fluctuations, whereas $C_{\Psi}$ measures the relative location of particles with respect to their nearest neighbors and is therefore immune to any large-scale fluctuations.\cite{Vivek17} It is worth of noting that although $C_{\Psi}$ is normally used to measure the bond-orientation order,\cite{Zahn00} its relaxation shows a quantitatively similar trend as the relaxation of the cage-relative translational correlation function in amorphous systems (Fig.~\ref{Figure8} in Appendix A). Thus, $C_{\Psi}$ has been generally used to characterize particle relaxations in the glass transition without the influence of long-wavelength fluctuations.\cite{Flenner15,Vivek17,Illing17} 

The decays of $F_s$ and $C_{\Psi}$ show similar relaxation times at low $\phi$ in absence of the long-wavelength fluctuations, but become significantly different at high $\phi$ near the glass transition (Fig.~\ref{Figure2}c). $F_s$ decays much faster due to the long-wavelength fluctuations.\cite{Vivek17,Illing17} We extract the $\alpha$-relaxation times of the two different correlation functions, $\tau_{F}$ and $\tau_{C}$, by fitting $F_s$ and $C_{\Psi}$ with stretched exponential functions $\sim \exp[-(t/\tau_{F,C})^\beta]$ at long times, where $\beta \leq 1$ is a stretching exponent. Smaller $\tau_{F}$ relative to $\tau_{C}$ indicates stronger fluctuations. Hence, the ratio $\tau_C/\tau_F$ provides a quantitative measurement on the strength of long-wavelength fluctuations as we shall justify below. 

The MSDs of particles at long times can be divided into two contributions. One is induced by the long-wavelength MW fluctuations indicated by $\langle r^2 \rangle_{MW}$; the other is due to the local structural relaxation of supercooled liquids indicated by $\langle r^2 \rangle_i$. Assume the two contributions are independent,\cite{Shiba18} we have
\begin{equation}
\langle r^2 \rangle = \langle r^2 \rangle_{MW} + \langle r^2 \rangle_i.
\end{equation}   
Therefore, the strength of the MW fluctuations can be written as
\begin{equation}
\langle r^2 \rangle_{MW} = \langle r^2 \rangle - \langle r^2 \rangle_i = Dt - D_i t, \label{MWs}
\end{equation}   
where in the last step we use the fact that MSDs become diffusive at long times for supercooled liquids, i.e., $\langle r^2 \rangle =Dt$ and $\langle r^2 \rangle_i=D_i t$ when $t$ is large. The total MSD is measured by the self-intermediate scattering function $F_s$. So we have $D \approx d_s^2/\tau_F$. The local structural relaxation without the influence of the long-wavelength fluctuations is characterized by the bond-orientation correlation function $C_{\Psi}$. Thus, we have $D_i \approx (\alpha d_s)^2/\tau_C$, where $\alpha < 1$ reflecting the fact that a particle only needs to move a small fraction of its own diameter to significantly change its relative orientation with respect to its nearest neighbors. Replacing $D$ and $D_i$ in Eq.~\ref{MWs} and setting $t = \tau_C$, the longest relaxation time of all processes, we have  
\begin{equation}
\langle r^2 \rangle_{MW} = d_s^2\left(\frac{\tau_c}{\tau_F} - \alpha^2\right). 
\end{equation}   
Hence, the strength of the MW fluctuations is linearly related to $\tau_C/\tau_F$. When $\alpha^2 \ll \tau_C/\tau_F$, a condition that is satisfied in our experiments, the strength of the fluctuations should be directly proportional to $\tau_C/\tau_F$. 

We plot $\tau_{C}/\tau_{F}$ against the size of systems (Fig.~\ref{Figure2}d). A logarithmic increase of $\tau_{C}/\tau_{F}$ with $R/d_s$ is again observed at high $\phi$. The results directly verify the logarithmic dependence of the strength of the long-wavelength fluctuations on system sizes in quasi-2D supercooled fluids close to the glass transition. At low $\phi$ where the long-wavelength fluctuations are weak, $\tau_{C}/\tau_{F}$ saturates for large systems (Fig.~\ref{Figure6}b in Appendix A), similar to the behavior of $\langle r^2 \rangle_{P}$. Qualitative similar results have also been observed for confined systems of the low-aspect-ratio ellipsoids (Fig.~\ref{Figure9} in Appendix A). In combination, our measurements on $\langle r^2 \rangle_{P}$ and $\tau_{C}/\tau_{F}$ provide direct experimental evidence of the existence of the MW fluctuations in quasi-2D colloidal supercooled liquids.

\begin{figure}
	\begin{center}
		\includegraphics[width=3.5in]{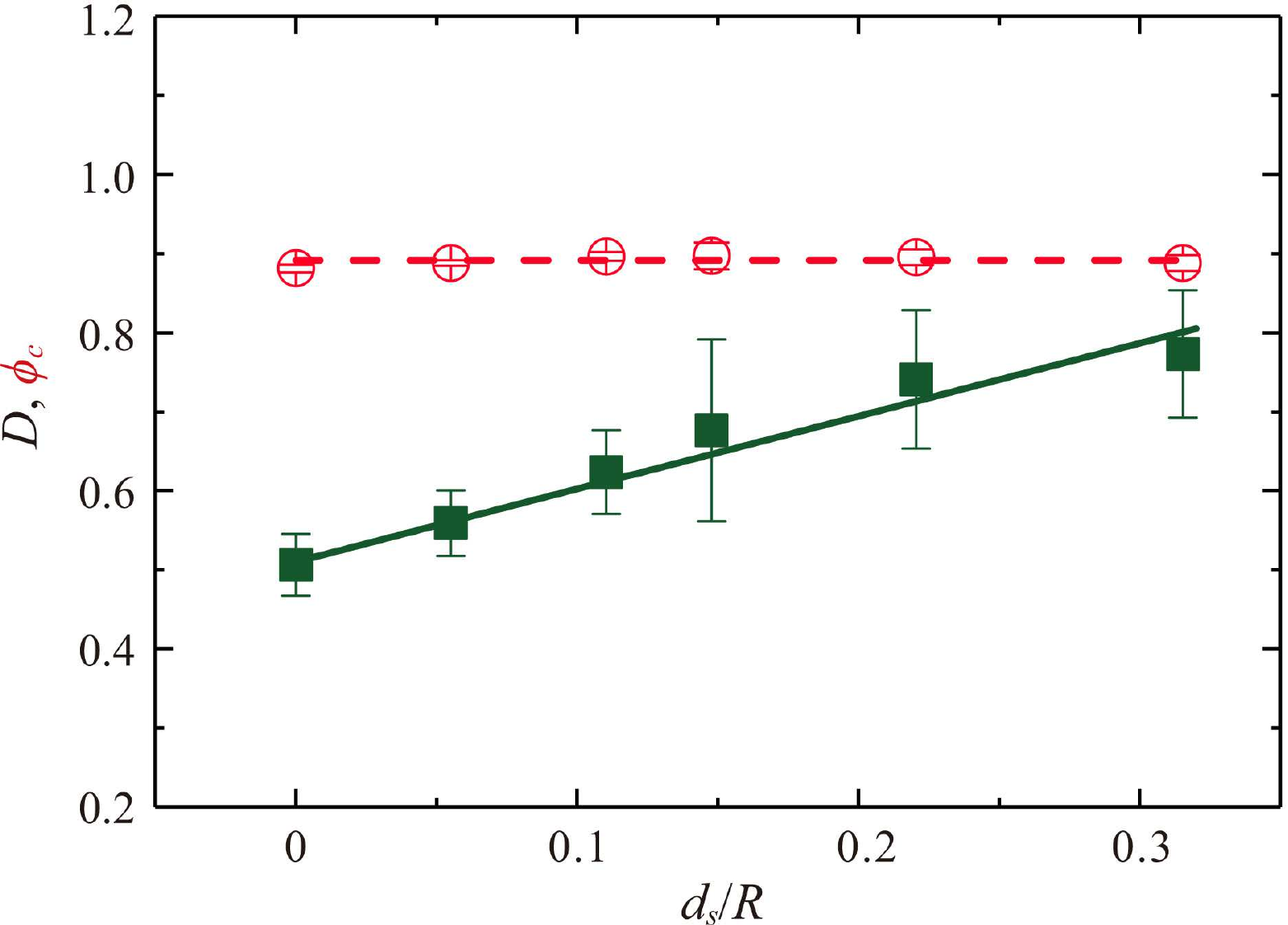}
	\end{center}
	\caption[Glass transition of 2D confined samples]
	{Relaxation of 2D confined supercooled liquids. The ideal glass transition area fraction $\phi_c$ (open red circles) and the fragility index $D$ (solid green squares) as a function of the inverse system size $d_s/R$. The dashed line indicates a constant $\phi_c=0.891$. The solid line is a linear fit with a slope $c=0.92$.
	}
	\label{Figure3}
\end{figure} 

\subsection{Static correlations in 2D glass transition}

Using confinement as a tool, we also measure the structural correlations of quasi-2D supercooled liquids. To avoid the influence of the MW fluctuations, we quantify the relaxation of supercooled liquids using $\tau_{C}$. $\tau_{C}$ as function of $\phi$ under different confinements can be fitted using the Vogel-Fulcher-Tammann (VFT) relation (Fig.~\ref{Figure10} in Appendix B), 
\begin{equation}
\tau_C = \tau_0 \exp\left[ \frac{D\phi}{\left(\phi_c-\phi\right)}\right], \label{VFT}
\end{equation} 
where $\tau_0$ is the microscopic time scale and $D$ is the fragility index. $\phi_c$ is the ideal glass transition area fraction, where $\tau_{C}$ diverges. Fig.~\ref{Figure3} shows $\phi_c$ and $D$ at different confinements for suspensions of spherical particles. Qualitatively similar to 3D supercooled liquids,\cite{Bo16} we find that $\phi_c$ is a constant, whereas $D$ decreases with $R$ following $D(R) = D(\infty) + c (d_s/R)$ with $c = 0.92 \pm 0.07$ and $D(\infty) = 0.51 \pm 0.04$. In both 2D and 3D, bulk samples show the most fragile behavior. At a fixed $\phi$, relaxation slows down drastically under confinement.\cite{Nugent07,Sarangapani08} A similar trend has also been observed for low-aspect-ratio ellipsoidal suspensions under confinement (Fig.~\ref{Figure11} in Appendix B). Hence, our experiments demonstrate that the nature of 2D and 3D glass transitions are qualitatively the same, a result corroborating recent studies.\cite{Vivek17,Illing17}

The constant $\phi_c$ and the linear increase of $D$ with $1/R$ allow us to extract a divergent structural correlation length, the so-called pinning length $\xi_p$,\cite{Charkrabarty15} in 2D supercooled colloidal liquids: From the Adam-Gibbs theory,\cite{Gibbs65} the $\alpha$-relaxation time of supercooled liquids can be written as 
\begin{equation}
\tau_C = \tau_0 \exp\left(\frac{A_0\phi}{s_c}\right), \label{GibbsEq}
\end{equation} 
where $A_0$ is a constant and $s_c$ is the configurational entropy density. For bulk samples without confinement, $s_c$ decreases to zero linearly at the ideal glass transition point $\phi_c$, $s_c=k(\phi_c-\phi)$, where $k$ is another proportional constant. Thus, Eq.~\ref{GibbsEq} turns into the VFT relation (Eq.~\ref{VFT}) if we identify $D$ as $A_0/k$. For a confined sample of size $R$, Chakrabarty {\it et al.} proposed that the configurational entropy density follows 
$s_c(\phi,R)=F(R)s_c(\phi,\infty)$, where $F(R)$ is an increasing function of $R$ with $F(\infty) = 1$ and $F(0) = 0$.\cite{Charkrabarty15} Thus, the relaxation time of confined systems is 
\begin{equation}
\tau_C = \tau_0 \exp \left[ \frac{A_0\phi}{kF(R)(\phi_c-\phi)}\right], \label{GibbsNEW}
\end{equation}
which gives the system-size-dependent fragility index $D(R)=A_0/(kF(R))$. Since $F(R)$ is an increasing function of $R$, $D$ decreases with $R$, qualitatively agreeing with our experimental observation (Fig.~\ref{Figure3}). 
The ratio of the relaxation times between the confined sample and the bulk sample at a fixed $\phi$ can be calculated as
\begin{equation}
\ln\left[ \frac{\tau_c(\phi,R)}{\tau_c(\phi,\infty)}\right]=D(R)\left[ \frac{\phi}{\phi_c(R)-\phi}\right]-D(\infty)\left[ \frac{\phi}{\phi_c(\infty)-\phi}\right].  
\end{equation}
In general, $\phi_c$ could be a function of $R$. $D(\infty)$ and $\phi_c(\infty)$ indicate the fragility index and the ideal glass transition area fraction of the bulk sample, respectively. We then introduce the number density of pinned particles, i.e., the number of pinned particles divided by the total area of the system (ignoring any order one constant) as $\rho_{pin}=(R/d_s)/R^2 = 1/(Rd_s)$ in the above equation, which gives  
\begin{multline}
\ln\left[\frac{\tau_c(\phi,R)}{\tau_c(\phi,\infty)}\right] = \left[ D(R)\left( \frac{\phi}{\phi_c(R)-\phi}\right) \left(\frac{R}{d_s}\right) \right. \\ \left. - D(\infty)\left( \frac{\phi}{\phi_c(\infty)-\phi}\right) \left(\frac{R}{d_s}\right)\right] d_s^2\rho_{pin}.  \label{Ratio}
\end{multline}
$\rho_{pin}$ is a control parameter in the RFOT theory to reveal the structural correlation of glass transition.\cite{Cammarota12,Chakrabarty16} Here, we have extended the definition of $\rho_{pin}$ for random pinning in theory \cite{Cammarota12,Charkrabarty15} to pinning walls in our experiments. Numerical simulations have demonstrated that the growth of static correlations is qualitatively similar under these different pinning geometries.\cite{Berthier12}   
Since $\rho_{pin}$ has a dimension of [length]$^{-2}$, the prefactor in Eq.~\ref{Ratio} must has a dimension of [length]$^2$, which for convenience we can call as $\xi_p^2$, i.e.,
\begin{equation}
\ln \left[ \frac{\tau_c(\phi,R)}{\tau_c(\phi,\infty)}\right]  =  \xi_p^2 \rho_{pin} \label{fitting} 
\end{equation}
\begin{equation}
\xi_p  \equiv  \left[ D(R)\left( \frac{\phi}{\phi_c(R)-\phi}\right) \left(\frac{R}{d_s}\right) - D(\infty)\left( \frac{\phi}{\phi_c(\infty)-\phi}\right) \left(\frac{R}{d_s}\right) \right]^{1/2} d_s. \label{Ratio2} 
\end{equation}
Next, we apply our experimental observations. First, our measurements show that $\phi_c$ is independent of $R$ (Fig.~\ref{Figure3}). Thus, Eq.~\ref{Ratio2} becomes 
\begin{equation}
\xi_p(\phi,R)=\left[ [D(R)-D(\infty)] \left(\frac{R}{d_s}\right)\right]^{1/2} \left(\frac{\phi}{\phi_c-\phi}\right)^{1/2} d_s.
\end{equation}
Furthermore, our experiments show that $D$ increases linearly with the inverse system size $d_s/R$ (Fig.~\ref{Figure3}). Thus, the term in the first bracket on the right is a constant and equals to the slope of the linear relation, $c = 0.92 \pm 0.07$. Therefore, we finally reach 
\begin{equation}
\xi_p(\phi)=c^{1/2}\left( \frac{\phi}{\phi_c-\phi} \right)^{1/2} d_s, \label{length} 
\end{equation}                                            
which is independent of system size $R$. Such a length can be taken as the static structural correlation length of bulk supercooled liquids. It should be emphasized that any valid static structural correlation length of glass transition must be a bulk property of supercooled liquids, independent of the size of confinement. The RFOT theory uses confinement as a tool to probe the existence of structural correlations in bulk samples.\cite{Kirkpatrick89,Biroli08,Cammarota12,Chakrabarty16}

\begin{figure}
	\begin{center}
		\includegraphics[width=3.5in]{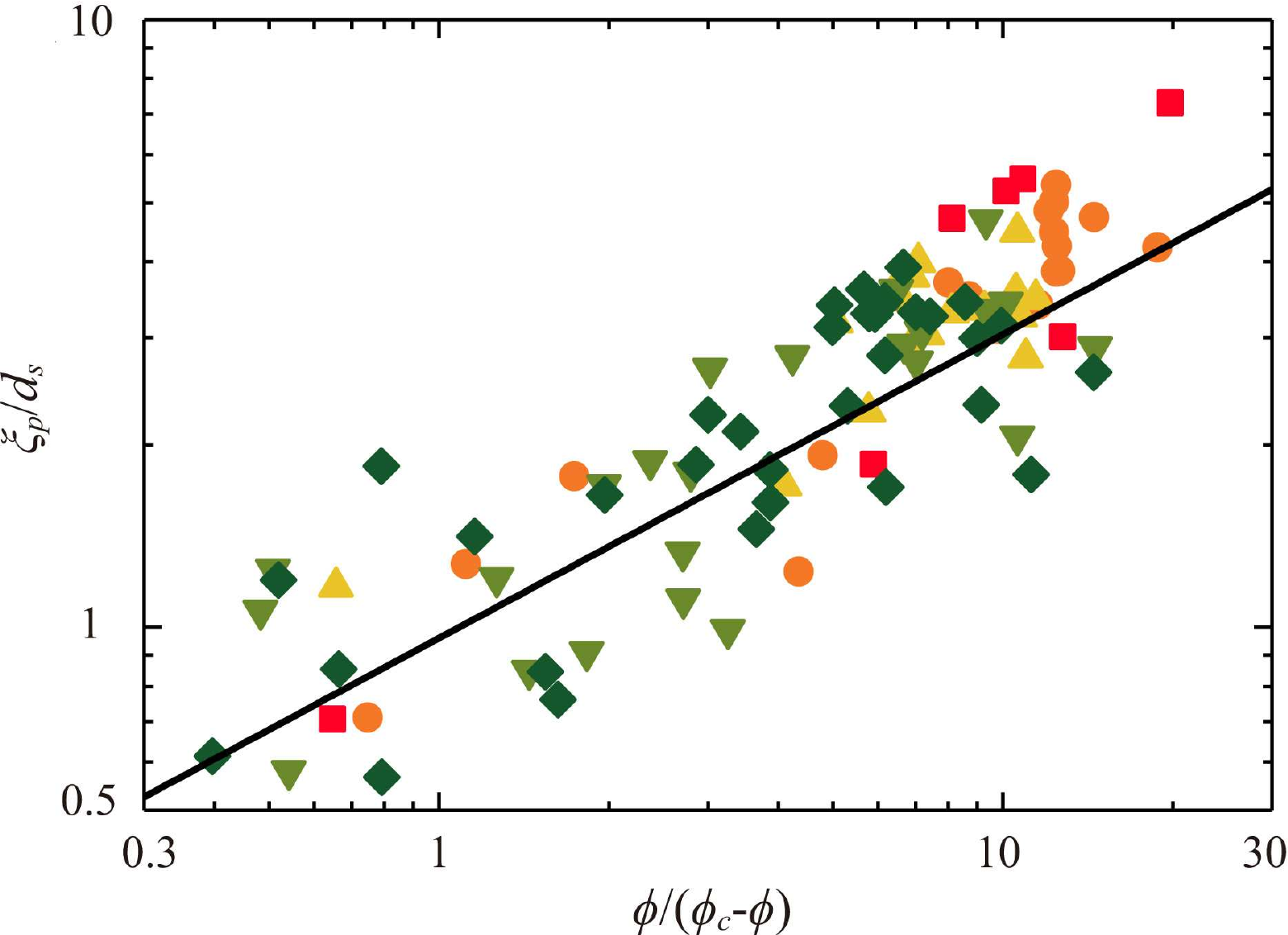}
	\end{center}
	\caption[Structure correlation length]
	{Structural correlation length of 2D supercooled colloidal liquids. The pinning length $\xi_p/d_s$ versus the rescaled area fraction $\phi/(\phi_c-\phi)$. Different symbols are from different $R$. The solid line is $\xi_p/d_s=[c\phi/(\phi_c-\phi)]^{1/2}$.
	}
	\label{Figure4}
\end{figure} 

Eq.~\ref{length} shows that $\xi_p$ diverges at $\phi_c$ following a square-root scaling. Our direct measurements of $\xi_p$ are indeed consistent with this prediction (Fig.~\ref{Figure4}). Here, to extract $\xi_p$ directly from experiments, we try to collapse all experimental relaxation times $\tau_c(\phi,R)/\tau_c(\phi,\infty)$ at different $\phi$ and $R$ as a function of $\rho_{pin}$ into a master curve by manually adjusting a single fitting parameter, $\xi_p(\phi,R)$ (Eq.~\ref{fitting}). The resulting $\xi_p$ is independent of $R$ and shows a power-law scaling consistent with Eq.~\ref{length} (Fig.~\ref{Figure4}). In comparison, $\xi_p/d_s \sim \lbrack\phi/\left(\phi_c-\phi\right)\rbrack^{1/3}$ in 3D.\cite{Bo16}

The results in both 2D and 3D samples support the hypothesis that the configurational entropy of confined systems decreases as $s_c(\phi,R) = F(R)s_c(\phi,\infty)$,\cite{Charkrabarty15,Chakrabarty16} where $F(R) \sim 1/D(R)$ is an increasing function of $R$ between zero and one. Notice that the point-to-set correlation length $\xi \sim [\phi/(\phi_c-\phi)]^{1/(d-\theta)}$ from Eq.~(\ref{equ1}) when we replace $Ts_c=Tk(T-T_K)$ with $[k(\phi_c-\phi)]/\phi$ suitable for colloidal suspensions.\cite{Gokhale16,Weeks17} Thus, our pinning length should scale with $\xi$ in $d$ dimensions as 
\begin{equation}
\xi_p \sim [\phi/(\phi_c-\phi)]^{1/d} \sim \xi^{(d-\theta)/d}, \label{power} 
\end{equation} 
quantitatively the same as the scaling of a similar pinning length predicted by a RFOT theory.\cite{Cammarota12} However, in Ref.~14, a different formula of the configurational entropy is assumed, where $s_c$ decrease with increasing confinement following $s_c(\phi,R) = s_c(\phi,\infty) - f(\phi)/R$. Here, $f$ is a function of $\phi$ and independent of $R$. It is worth of checking if such a discrepancy can be explained within the framework of RFOT \cite{Chakrabarty16} or if alternative theories such as dynamical facilitation theories,\cite{Chandler10} geometric frustration theories \cite{Tarjus05} and models with short-ranged orientational orders \cite{Watanabe11,Russo15} are needed.     

It has been shown within RFOT, $\xi_p \sim \xi^{1/d}$ near the random first order transition and $\xi_p \sim \xi$ away from the transition.\cite{Charbonneau12} Thus, Eq.~\ref{power} from our experiments indicates $\theta = d-1$ near the transition and $\theta = 0$ away from the transition.   

Lastly, it is worth of noting that the confining boundary adopted in our study is smooth and different from the frozen equilibrium boundary suggested by RFOT.\cite{Biroli08,Nagamanasa15} Nevertheless, a recent RFOT theory has shown that confined systems with smooth boundaries exhibit qualitatively similar trends in the development of static correlations and, therefore, can be used as a tool to probe static correlation lengths in supercooled liquids.\cite{Cammarota13} It would be interesting to check how the relaxation dynamics of particles change when different confining boundary conditions are applied in experiments.    

\subsection{The effect of long-wavelength fluctuations on translational and orientational relaxations of ellipsoids} 

Finally, after confirming the existence of the MW fluctuations and exploring the static correlations in 2D confined systems, we set out to investigate how the MW fluctuations affect the glass transition of anisotropic particles in 2D. Although most works on the translational and orientational relaxations of supercooled liquids of anisotropic particles focus on 2D systems, \cite{Schreck10,Shen12,Zheng11,Zheng14,Vivek17_2} very few studies have explored the possible differential influences of the MW fluctuations on the translational and orientational degrees of freedom of the systems. Here, we probe the effect of the MW fluctuations on the translation and orientation of quasi-2D colloidal suspensions of ellipsoids and re-exam the established results on the translational and orientational relaxations of 2D anisotropic supercooled liquids.  

To quantify particle orientational dynamics, we measure the body orientational correlation function, $L_n(t)$ \cite{Zheng11}
\begin{equation}
L_n(t) = \left\langle \frac{1}{N} \sum_{j} \cos\left[n\left(\alpha_j(t+t_0)-\alpha_j(t_0)\right)\right] \right\rangle_{t_0}, \label{body-orientation}
\end{equation} 
where $\alpha_j(t)$ is the orientation of the major axis of ellipsoid $j$ with respect to the $x$ axis. We fix $n = 6$ in our analysis. Different choices of $n$ yield qualitatively similar results (Fig.~\ref{Figure12} in Appendix C).\cite{Zheng11,Zheng14} It should be emphasized that the body orientational correlation of anisotropic particles defined in Eq.~\ref{body-orientation} is different from the bond orientational correlation defined in Eq.~\ref{Cpsi} that exists even for spherical particles in 2D as shown by the KTHNY theory and discussed more recently in the context of glass transition.\cite{Flenner19} The body orientational relaxation time at a given $\phi$, $\tau_L(\phi)$, can again be extracted by fitting $L_n(t)$ with stretched exponential functions at long times.   

\begin{figure}[t]
	\begin{center}
		\includegraphics[width=3.0in]{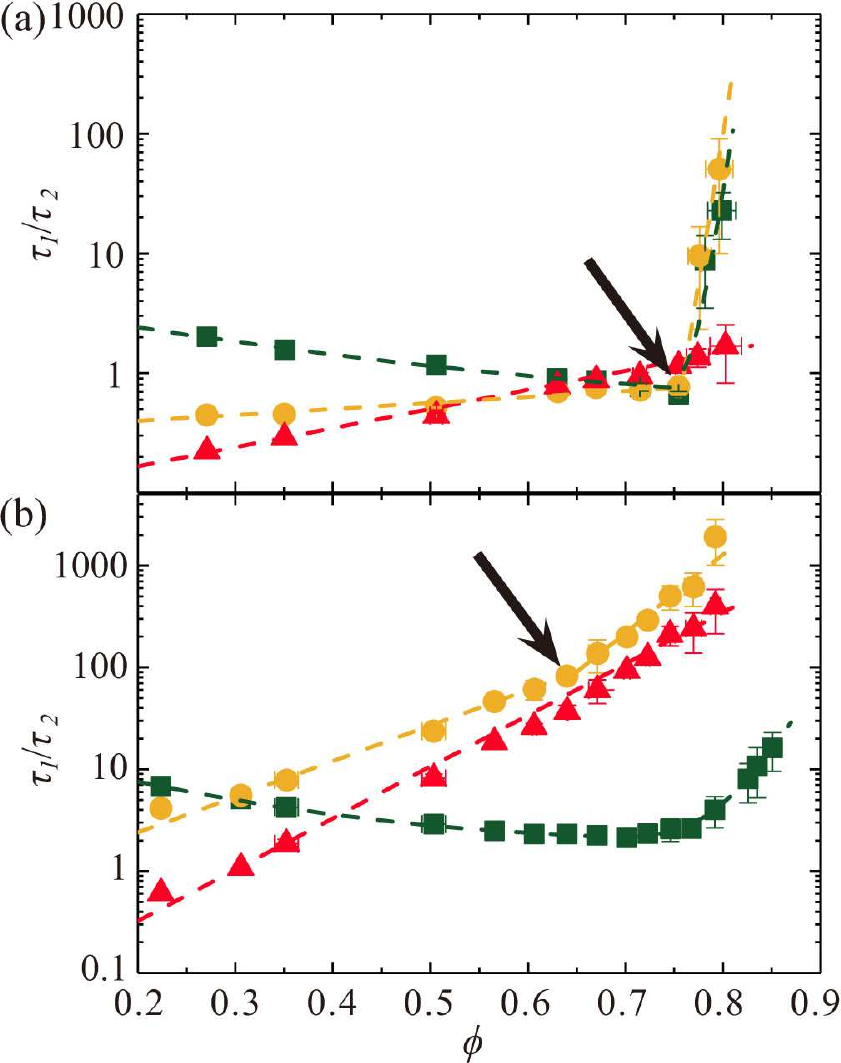}
	\end{center}
	\caption[Influence of Mermin-Wagner fluctuations on translational and diffusional dynamics of ellipsoids. ]
	{Influence of the Mermin-Wagner fluctuations on translational and orientational relaxations of ellipsoids. The aspect ratios of ellipsoids are $P=1.9$ (a) and 6.7 (b). There different relaxation time ratios versus area fraction:  $\tau_{C}/\tau_{F}$ (green squares), $\tau_{L}/\tau_{F}$ (yellow disks), and $\tau_{L}/\tau_{C} $ (red triangles). Yellow and green dashed lines are visual guidelines. Red dashed lines are linear fits in the log-linear scale. Black arrows indicate the kinks in $\tau_L/\tau_F$ versus $\phi$, marking the onset of the orientational glass phase. 
	}
	\label{Figure5}
\end{figure}

By comparing $\tau_F$ from $F_s$ with $\tau_L$ from $L_n$, Zheng {\it et al.} identified a two-step glass transition for quasi-2D suspensions of ellipsoids,\cite{Zheng11,Zheng14} where the glass transition of particles' orientational degree of freedom occurs at a lower $\phi$ than the glass transition of the translation of particles. A phase of orientational glasses emerges when the area fraction of a sample is larger than the orientational glass transition area fraction $\phi_{g}^{o}$ but smaller than the translational glass transition area fraction $\phi_{g}^{t}$.\cite{Letz00,Zheng11} The rotational relaxation of an orientational glass is much slower than its translational relaxation. The density difference $\Delta \phi = \phi_{g}^{o} - \phi_{g}^{r}$ decreases with decreasing $P$ and vanishes when $P = 1.7$ for ellipsoids with hard-sphere potentials. Since $\tau_F$ is significantly affected by the MW fluctuations in 2D supercooled liquids (Figs.~\ref{Figure2}c and d), the nature of the two-step glass transition and the orientational glasses may strongly depend on the MW fluctuations.     

To demonstrate the influence of the MW fluctuations on the two-step glass transition of anisotropic particles, we plot three different relaxation time ratios: $\tau_C/\tau_F$, $\tau_L/\tau_F$ and $\tau_L/\tau_C$ for bulk samples at different $\phi$ (Fig.~\ref{Figure5}). Here, $\tau_C/\tau_F$ indicates the strength of the MW fluctuations (Fig.~\ref{Figure2}d). $\tau_L/\tau_F$ captures the two-step glass transition discussed in Ref.~12, whereas $\tau_L/\tau_C$ compares the orientational and transitional relaxations after the long-wavelength fluctuations are removed. For ellipsoids of both small and large $P$, the MW fluctuations emerge at high $\phi$ with their strength increasing with $\phi$ (green squares in Fig.~\ref{Figure5}). At high $\phi$ where the fluctuations are strong, a qualitative change in the trend of $\tau_L/\tau_F$ versus $\phi$ can be observed: the increase of $\tau_L/\tau_F$ becomes significantly faster above a threshold $\phi$, manifesting as upturning kinks in the curves (yellow disks in Fig.~\ref{Figure5}). The effect is most pronounced for the low-$P$ samples. Such qualitative changes arise from a sharp increase of $\tau_L$ relative to $\tau_F$, signaling the formation of orientational glasses.\cite{Letz00,Zheng11, Zheng14} Since the regime of orientational glasses $\Delta \phi$ is small for ellipsoids of small $P$, the increase of $\tau_L/\tau_F$ is sharper over a narrower range of $\phi$ for the low-$P$ samples (Fig.~\ref{Figure5}a). Importantly, when the MW fluctuations are removed, $\tau_L/\tau_C$ shows a smooth exponential increase over the whole range of $\phi$ of our experiments without any qualitative change in trend for both low and high-$P$ samples (red triangles in Fig.~\ref{Figure5}). This Arrhenius increase of $\tau_L/\tau_C$ suggests a constant energy-barrier difference between the translational and orientational relaxations. The observation furthermore suggests that the two-step glass transition and the orientational glasses in 2D are the consequence of the MW fluctuations. Translational and orientational relaxations are qualitatively similar at local scales.

\section{Conclusions}
We experimentally studied the long-wavelength fluctuations and structural correlations in quasi-2D suspensions of spherical and ellipsoidal particles. Circular confinement induced by optical traps was employed as a tool in our study to investigate two bulk properties of colloidal supercooled liquids. First, confinement was used to measure the dependence of long-wavelength thermal fluctuations on the linear size of systems. Our experiments provided direct evidence of the existence of the Mermin-Wagner fluctuations in 2D supercooled liquids close to the glass transition. Second, using confinement as a probe, we also illustrated the emergence of structural correlations in 2D supercooled liquids near the glass transition. We extracted the corresponding static correlation length, which follows a divergent power-law scaling, qualitatively similar to that of 3D systems. Furthermore, we also demonstrated the differential influences of the Mermin-Wagner fluctuations on the translational and orientational relaxations of anisotropic particles. The relaxations of these two degrees of freedom show a qualitative similar trend at local scales, where the long-wavelength fluctuations are insconsequential. Taken together, our experimental study provided new insights not only on the structures and dynamics of 2D supercooled liquids, but also on the role of dimensionality on the glass transition. While the dimensionality-dependent long-wavelength thermal fluctuations strongly affect the dynamics of glass transition at large scales, microscopic structural relaxations are insensitive to these fluctuations. 

\section*{Conflicts of interest}
There are no conflicts to declare.

\section*{Acknowledgements}
We thank Yi Peng and Zhongyu Zheng for discussions and help with the experiments. The research was partially supported by the NSF MRSEC Program (DMR-1420013) and by the Packard Foundation.

\section*{Appendix A: Characterization of long-wavelength fluctuations}

\subsection*{A.1 Plateaued MSDs at the inflection point, $\langle r^2 \rangle_{P}$}

\begin{figure}
	\begin{center}
		\includegraphics[width=3in]{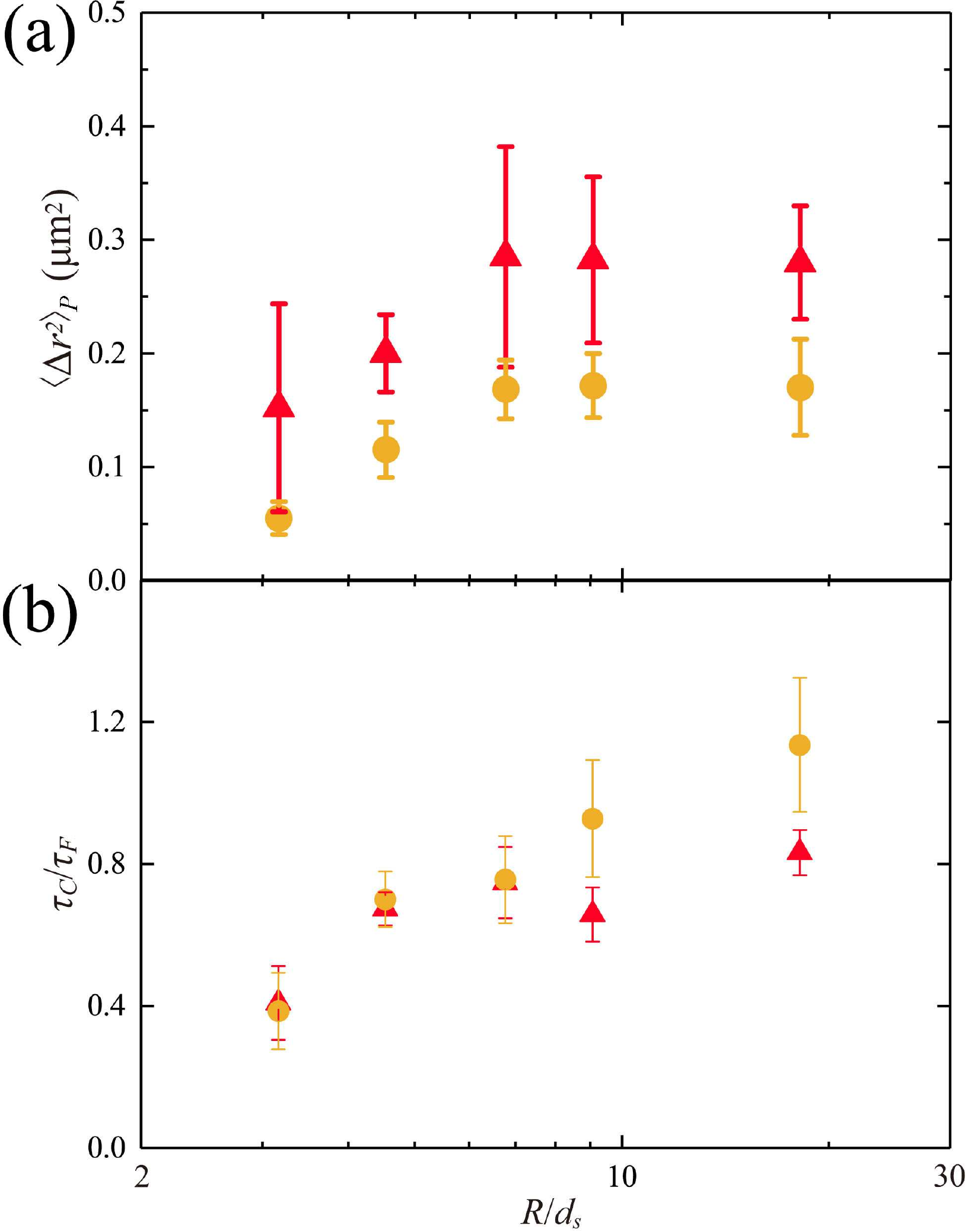}
	\end{center}
	\caption[Strength of the long-wavelength fluctuations]
	{Strength of the long-wavelength fluctuations. (a) Plateaued MSDs, $\langle r^2 \rangle_{P}$, as a functions of system size $R$ for $\phi = 0.80 \pm 0.01$ (yellow disks) and $\phi = 0.75 \pm 0.01$ (red triangles). For bulk samples, $\langle r^2 \rangle_{P} = 0.17 \pm 0.06$ $\mu$m$^2$ for $\phi = 0.80$ and $0.51 \pm 0.07$ $\mu$m$^2$ for $\phi = 0.75$. (b) Ratio of relaxation times, $\tau_C/\tau_F$, as a function of $R$ for $\phi = 0.80 \pm 0.01$ (yellow disks) and $\phi = 0.75 \pm 0.01$ (red triangles). For bulk samples, $\tau_C/\tau_F = 4.87 \pm 1.09$ for $\phi = 0.80$ and $0.98 \pm 0.17$ for $\phi = 0.75$.
	}
	\label{Figure6}
\end{figure}

We quantify the strength of the long-wavelength fluctuations by measuring the plateaued mean square displacements (MSDs), where the caging effect is most pronounced. The same approach has been used in the numerical study of the Mermin-Wagner (MW) fluctuations.\cite{Shiba16,Illing17} We identify the plateaued MSDs at the inflection point with $d^2\langle r^2 \rangle/dt^2=0$. At high $\phi = 0.82$, the plateaued MSDs, $\langle r^2 \rangle_{P}$, exhibits a logarithmic increase with the system size, the key signature of the MW fluctuations (Fig.~\ref{Figure2}b). At low $\phi$, when colloidal suspensions gradually transition from dense supercooled fluids to low-density fluids, the fluctuations should disappear. Interestingly, we find that when lowering $\phi$, the signature of the MW fluctuations diminishes for large systems first but persists in small systems. Specifically, $\langle r^2 \rangle_{P}$ shows the logarithmic increase at small $R$. The increase shows the trend of saturation at large $R$ (Fig.~\ref{Figure6}a). It is worth noting that for systems with large $R$, the plateau of MSDs becomes less distinct.\cite{Flenner15} The crossover from the short-time diffusive regime and the long-time diffusive regime is smooth lacking a true flat plateau. Although $\langle r^2 \rangle_{P}$ is still well defined mathematically in this case, it may not serve as a good indicator for the long-wavelength fluctuations. For the highest $\phi = 0.82$ we study, the value of $\langle r^2 \rangle_{P}$ of the bulk sample still indicates the presence of saturation in the large $R$ limit, even though the logarithmic dependence is observed for all the confined samples (Fig.~\ref{Figure2}b).

\subsection*{A.2 Bond-orientational correlation function, $C_{\Psi_n}$ vs. $C_{\Psi_6}$}

\begin{figure}
	\begin{center}
		\includegraphics[width=3in]{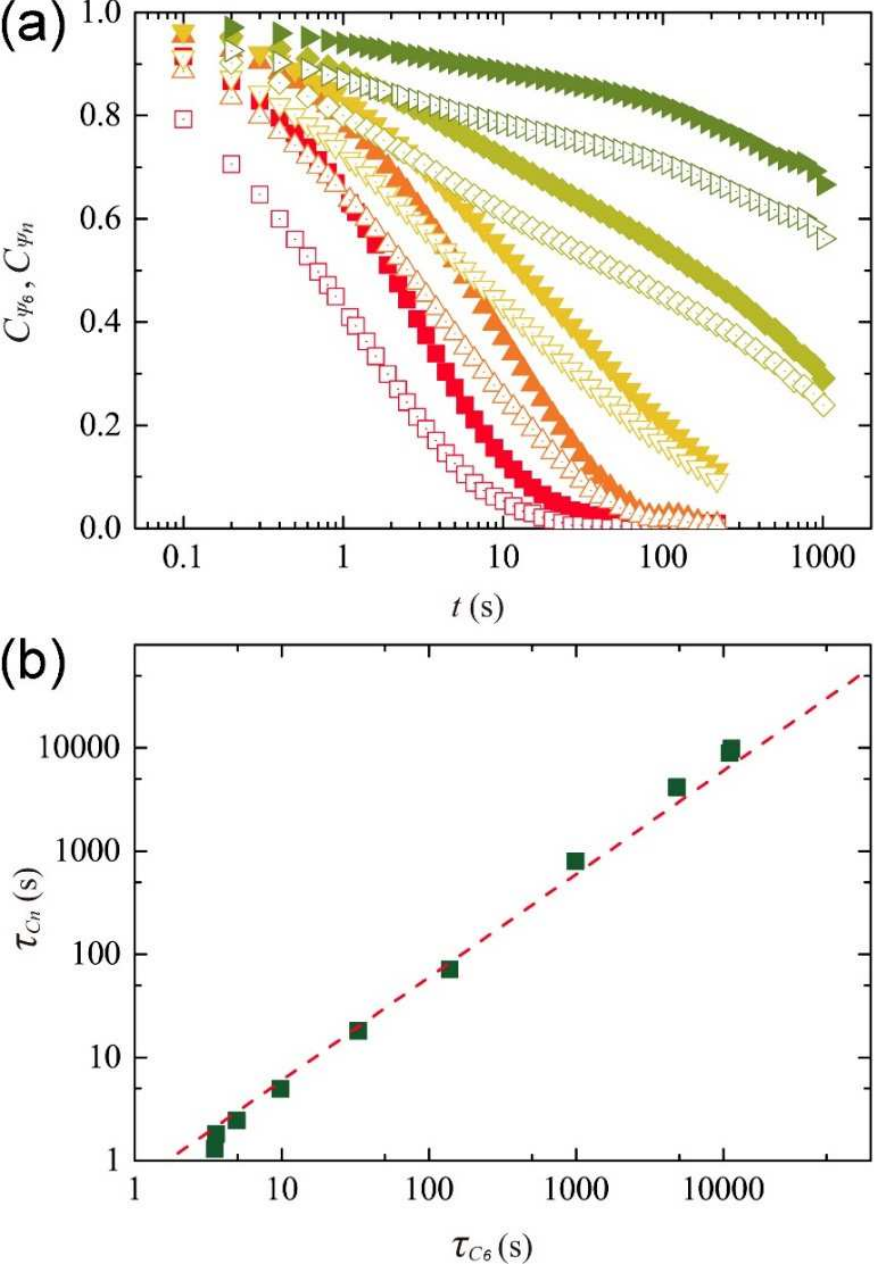}
	\end{center}
	\caption[Different definitions of the bond-orientational correlation functions]
	{Different definitions of the bond-orientational correlation functions. (a) Comparison of the six-fold bond-orientational correlation function with $n = 6$, $C_{\Psi_6}$, and the average bond-orientational correlation function with $n$ ranging from 4 to 8, $C_{\Psi_n}$. Solid symbols are for $C_{\Psi_6}$. Empty symbols are for $C_{\Psi_n}$. From left to right, the area fractions are $\phi = 0.229$, $0.695$, $0.776$, $0.814$ and $0.838$. (b) Relaxation time of $C_{\Psi_n}$, $\tau_{C_n}$, versus relaxation time of $C_{\Psi_6}$, $\tau_{C_6}$. The dashed line indicates $\tau_{C_n}=0.6\tau_{C_6}$.   
	}
	\label{Figure7}
\end{figure}

In the main text, we fix $n = 6$ in the definition of the bond-orientational correlation functions, $C_\Psi$ (Eq.~\ref{Cpsi}). In amorphous samples, particles may not have six nearest neighbors. Thus, we also calculate the bond-orientational correlation by averaging $n$ from 4 to 8 to include the majority of particles with different numbers of nearest neighbors:\cite{Illing17}
\begin{equation}
C_{\Psi}=\frac{\langle \sum_{n=4}^{8} \sum_{j}\left[\psi_n^j(t_0)\right]^*\cdot\left[\psi_n^j(t_0+t)\right]\rangle_{t_0}}{\langle \sum_{n=4}^{8}\sum_{j}|\psi_n^j(t_0)|^2\rangle_{t_0}}. 
\end{equation}
The six-fold configuration is a locally favored structure, which is therefore more relaxed and shows slower relaxations. As a result, $C_{\Psi_n}$ with $n$ ranging from 4 to 8 relaxes faster than $C_{\Psi_6}$ with $n$ fixed at 6 (Fig.~\ref{Figure7}a). Nevertheless, the relaxation times extracted from these two different correlation functions are linearly proportional to each other (Fig.~\ref{Figure7}b). Hence, the conclusion of our paper should not depend on the choice of these two different bond-orientational correlation functions.   

\subsection*{A.3 Local structural relaxation, $C_{\Psi}$ vs. $F_{s-CR}$}

\begin{figure}
	\begin{center}
		\includegraphics[width=3in]{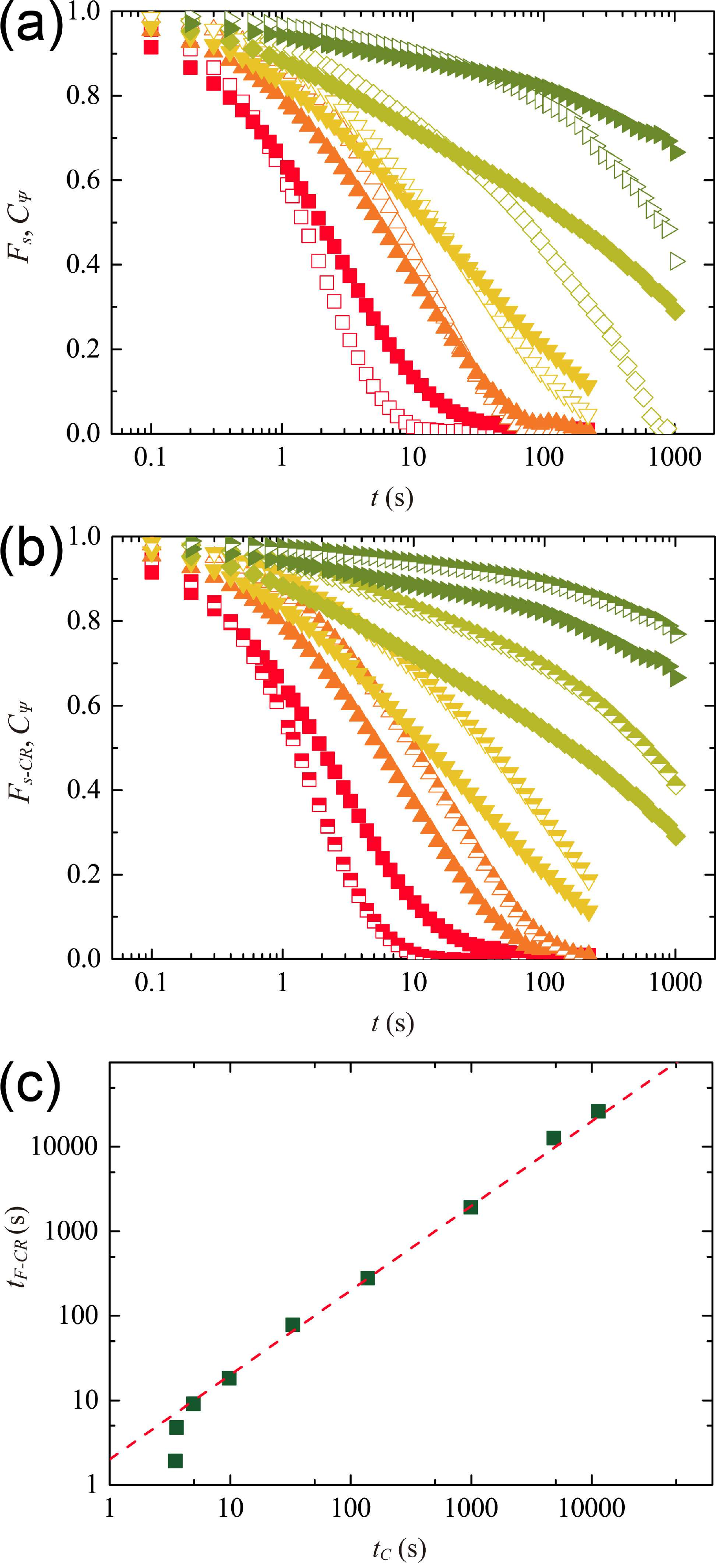}
	\end{center}
	\caption[Comparison of three correlation functions]
	{Comparison of three correlation functions, $F_s$, $F_{s-CR}$ and $C_\Psi$. (a) $F_s$ vs. $C_\Psi$ at different area fractions. From left to right, the area fraction $\phi = 0.229$, $0.695$, $0.776$, $0.814$ and $0.838$. Empty symbols are for $F_s$. Solid symbols are for $C_\Psi$. (b) $F_{s-CR}$ vs. $C_\Psi$. The area fractions are the same as those shown in (a). Half-empty symbols are for $F_{s-CR}$, which show similar relaxation as $C_\Psi$. (c) The relaxation time extracted from $F_{s-CR}$, $\tau_{F-CR}$, as a function of the relaxation time extracted from $C_\Psi$, $\tau_C$. The relaxation times are obtained by fitting the long-time decay of the corresponding correlation functions with stretched exponential functions. The dashed line indicates $\tau_{F-CR} = 2\tau_C$.   
	}
	\label{Figure8}
\end{figure} 

We quantify the local structural relaxation of supercooled colloidal liquids without the influence of long wavelength fluctuations using $C_{\Psi}$. The local structural relaxation can also be quantified by the cage-relative translational correlation function, $F_{s-CR}$, which is defined as\cite{Vivek17,Illing17} 
\begin{equation}
F_{s-CR}(t) = \langle \frac{1}{N} \sum_{j} \cos\left(\vec{Q}\cdot\Delta\vec{r}_{CR,j}(t_0,t)\right)\rangle_{t_0}, 
\end{equation}
where $\Delta\vec{r}_{CR,j}$ is the relative displacement of Particle $j$ with respect to its nearest neighbors. 
\begin{equation}
\Delta\vec{r}_{CR,j} = \vec{r}_j(t_0+t)-\vec{r}_j(t_0)-\frac{1}{M}\sum_{i}\left[\vec{r}_i(t_0+t)-\vec{r}_i(t_0)\right], 
\end{equation}
where $i$ denotes nearest neighbors of Particle $j$ at initial time $t_0$ and the sum is over all the $M$ neighbors. $F_{s-CR}$ is the counterpart of the self-immediate scattering function, $F_s$, after the removal of long-wavelength fluctuations. Refs. 8 and 9 showed that the relaxation of $F_{s-CR}$ and $C_\Psi$ follow a similar trend. We confirm this result in our experiments. Figure \ref{Figure8} shows $F_s$, $F_{s-CR}$ and $C_\Psi$ at different area fractions. At low area fractions, the three correlation functions exhibit similar relaxation trends. However, with increasing area fractions, the relaxation of $F_s$ becomes significantly faster due to the presence of long wavelength fluctuations in 2D systems (Fig.~\ref{Figure8}a). In contrast, without the influence of the fluctuations, $F_{s-CR}$ and $C_\Psi$ show similar relaxation behaviors (Fig.~\ref{Figure8}b). Quantitatively, the relaxation time extracted from $C_\Psi$, $\tau_C$, is linearly proportional to the relaxation times extracted from $F_{s-CR}$, $\tau_{F-CR}$, except at the lowest area fraction $\phi$ (Fig.~\ref{Figure8}c). Thus, our measurements verify that both $C_\Psi$ and $F_{s-CR}$ measure the same local structural relaxation without the influence of long-wavelength fluctuations.

\subsection*{A.4 Relaxation time ratio, $\tau_{C}/\tau_{F}$}

The logarithmic increase of $\tau_C/\tau_F$ for samples of spherical PMMA particles at high $\phi = 0.82$ indicates the logarithmic dependence of the strength of the long-wavelength fluctuations (Fig.~\ref{Figure2}d). For low $\phi$ samples, $\tau_C/\tau_F$ shows a similar trend as $\langle r^2 \rangle_{P}$ at low $\phi$ (Fig.~\ref{Figure6}b): For $\phi = 0.80$ samples, $\tau_C/\tau_F$ increases logarithmically for small systems and sub-logarithmically for large systems. For even lower $\phi = 0.75$ samples, $\tau_C/\tau_F$ increases logarithmically for small systems and saturates for large systems.
 
\begin{figure}
	\begin{center}
		\includegraphics[width=3in]{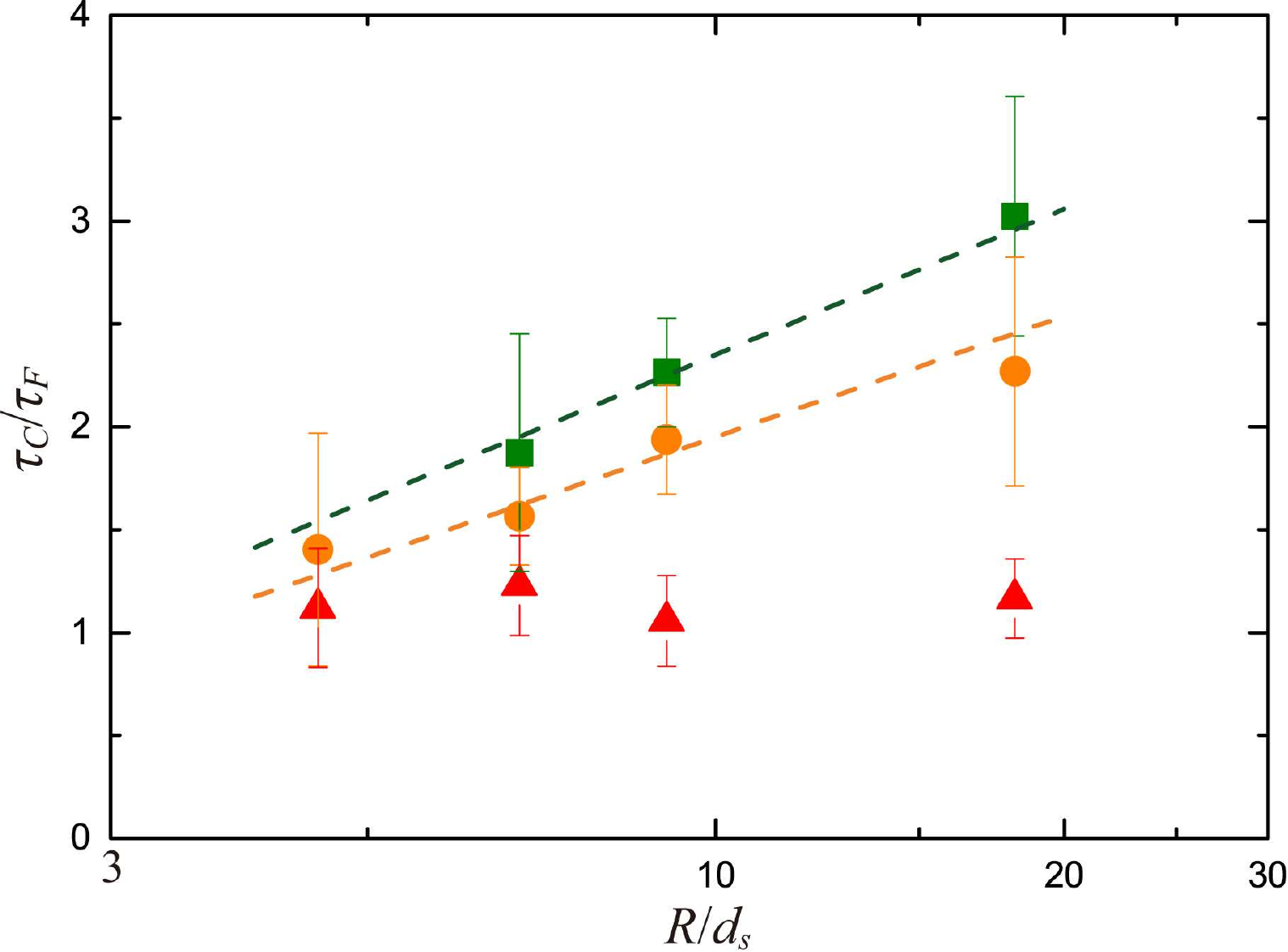}
	\end{center}
	\caption[Relaxation time of tau_C/tau_F]
	{Relaxation time ratio $\tau_C/\tau_F$ as a function of system size $R$ for the low-aspect-ratio ellipsoids with $P = 1.9$. From top to bottom, the area fraction is $\phi = 0.81$, $0.78$ and $0.75$. The dotted lines indicate the logarithmic dependence. $\tau_C/\tau_F = 19.2$ for bulk samples at $\phi = 0.81$.  
	}
	\label{Figure9}
\end{figure}

Similar behaviors have also been observed for systems composed of the low-aspect-ratio ellipsoids with $P = 1.9$ (Fig.~\ref{Figure9}). At high volume fractions, $\tau_C/\tau_F$ increases logarithmically with system size. At low volume fractions, it saturates when the system size is large. For the high-aspect-ratio ellipsoids with $P = 6.7$, we cannot pin the particles using optical tweezers, which prevents us from measuring fluctuations at different system sizes.

\section*{Appendix B: Structural correlations of quasi-2D supercooled liquids}

\subsection*{B.1 Relaxation of suspensions of spherical particles under confinement}

\begin{figure}
	\begin{center}
		\includegraphics[width=3in]{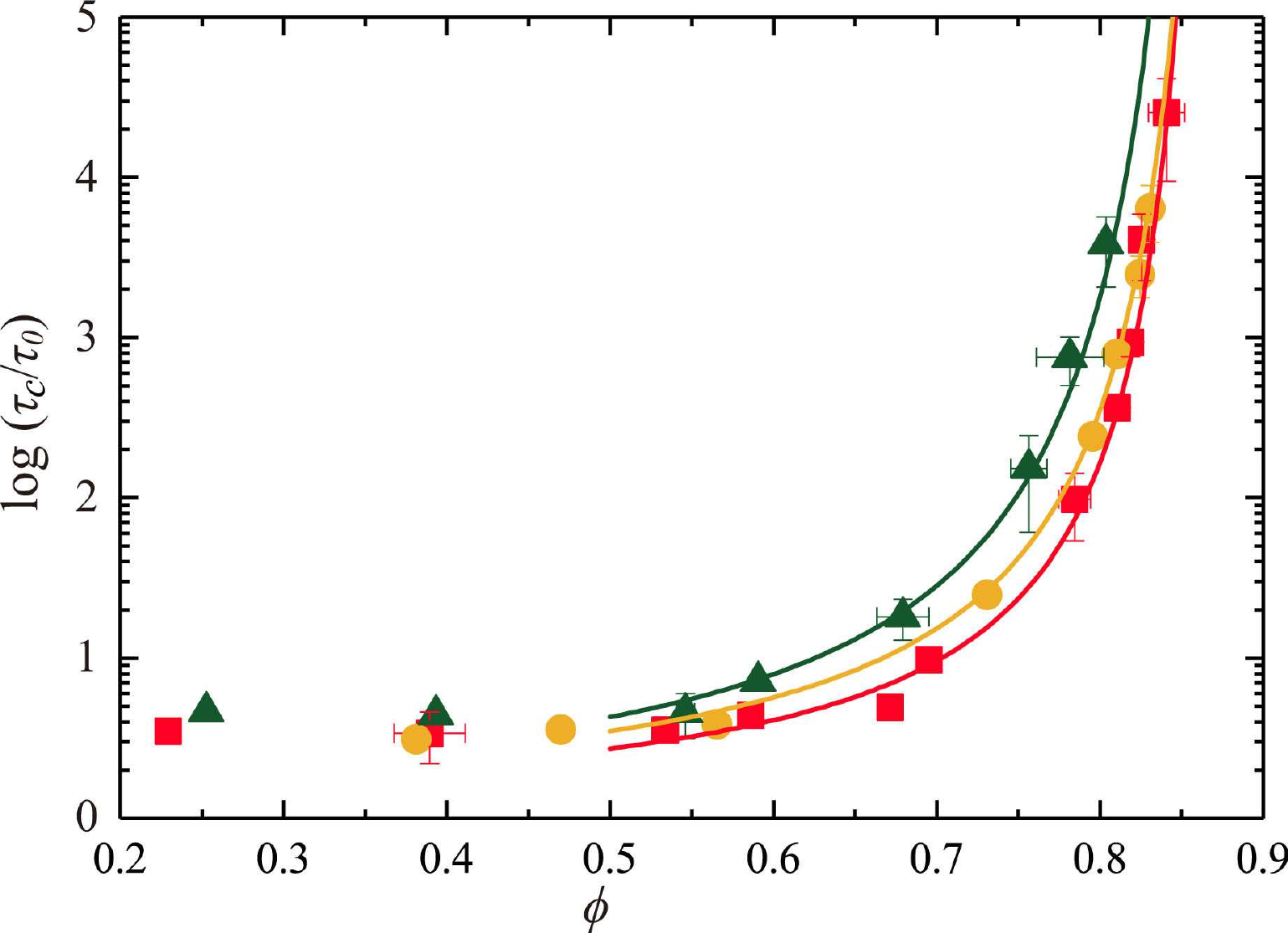}
	\end{center}
	\caption[Relaxation of suspensions of spherical particles]
	{Relaxation of suspensions of spherical particles. The $\alpha$-relaxation time $\tau_C$ as a function of the area fraction $\phi$ for different confinements. $\tau_C$ is extracted from the bond-orientational correlation function, $C_\Psi$. From top to bottom, $R = 3.2d_s$ (green triangles), $9.0d_s$ (yellow disks) and bulk samples (red squares). Six different confinements are measured in total (Fig.~\ref{Figure3}). We only show three here for clarity. $\tau_0 = 1.04$ s is the Brownian translational relaxation time of the small particles in the dilute limit. The solid lines are the VFT fittings. Only data close to the glass transition with $\phi > 0.5$ are used for fitting.   
	}
	\label{Figure10}
\end{figure}

We study the long-time relaxation of particle dynamics based on the bond-orientational correlation function, $C_{\Psi}$. The $\alpha$-relaxation time, $\tau_C$, is extracted by fitting $C_\Psi$ with an stretched exponential function at long times. Figure \ref{Figure10} shows $\tau_C(\phi)$ under different degrees of circular confinement. The relaxation of particles slows down dramatically as $\phi$ increases toward the glass transition. At a given $\phi$, the relaxation time increases with increasing confinement, similar to the behavior of 3D confined suspensions.\cite{Nugent07,Sarangapani08} We fit the super-Arrhenius increase of $\tau_C(\phi)$ near the glass transition with the classical Vogel-Fulcher-Tammann (VFT) relation (Eq.~\ref{VFT}) for each confinement (Fig.~\ref{Figure10}). $D$ gives the fragility index quantifying the local slope of $\log \tau_C(\phi)$ near the glass transition and $\phi_c$ represents the ideal glass transition area fraction where $\tau_C$ diverges. $D(R)$ and $\phi_c(R)$ for suspensions of spherical particles are shown in Fig.~\ref{Figure3}. 

\subsection*{B.2 Relaxation of suspensions of ellipsoids under confinement}

\begin{figure}
	\begin{center}
		\includegraphics[width=3in]{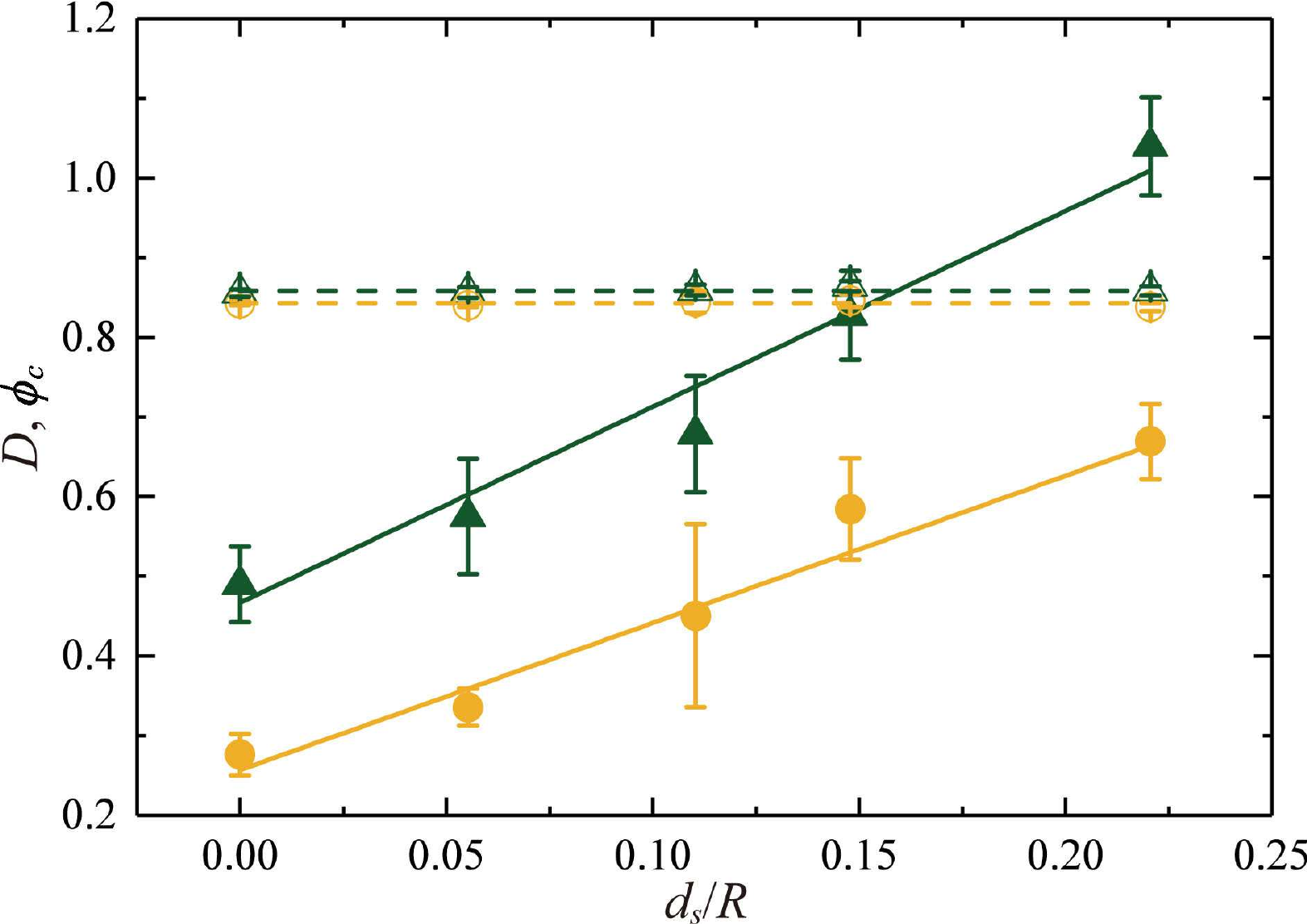}
	\end{center}
	\caption[Translational and orientational relaxations of suspension of ellipsoids]
	{Translational and orientational relaxations of suspension of ellipsoids with aspect ratio $P = 1.9$. $\phi_c$ (open symbols) and $D$ (solid symbols) from $C_{\Psi}$ (orange disks) and $L_n$ (green triangles) as a function of the inverse system size $d_s/R$. The dashed lines indicate constant values. The solid lines are linear fittings for $D$. 
	}
	\label{Figure11}
\end{figure}

We also measure the translational and orientational relaxations of suspensions made of ellipsoids of aspect ratio $P = 1.9$. The translational relaxation is again characterized by $\tau_C$ based on $C_\Psi$, whereas the orientational relaxation is quantified by $\tau_L$ based on the body orientational correlation function, $L_n$. Figure \ref{Figure11} shows the resulting $D(R)$ and $\phi_c(R)$, which exhibit qualitatively similar trends as those of spherical suspensions. For both the translational and orientational degrees of freedom, $\phi_c(R)$ is a constant independent of $R$. In both cases, $D$ increases linearly with the inverse system size. The slope of the linear relation is slightly larger for the orientational relaxation. The results also indicate the translational and orientational relaxations are qualitatively similar at local scales excluding the long-wavelength fluctuations. As noted above, we cannot trap ellipsoids with $P = 6.7$ using optical tweezers. Therefore, we cannot probe the relaxation of the high-aspect-ratio ellipsoids under confinement in our experiments.

\section*{Appendix C: Effect of long-wavelength fluctuations on translational and orientational relaxations of ellipsoids}

\subsection*{C.1 Body orientational correlation function, $L_n$}

\begin{figure}
	\begin{center}
		\includegraphics[width=3in]{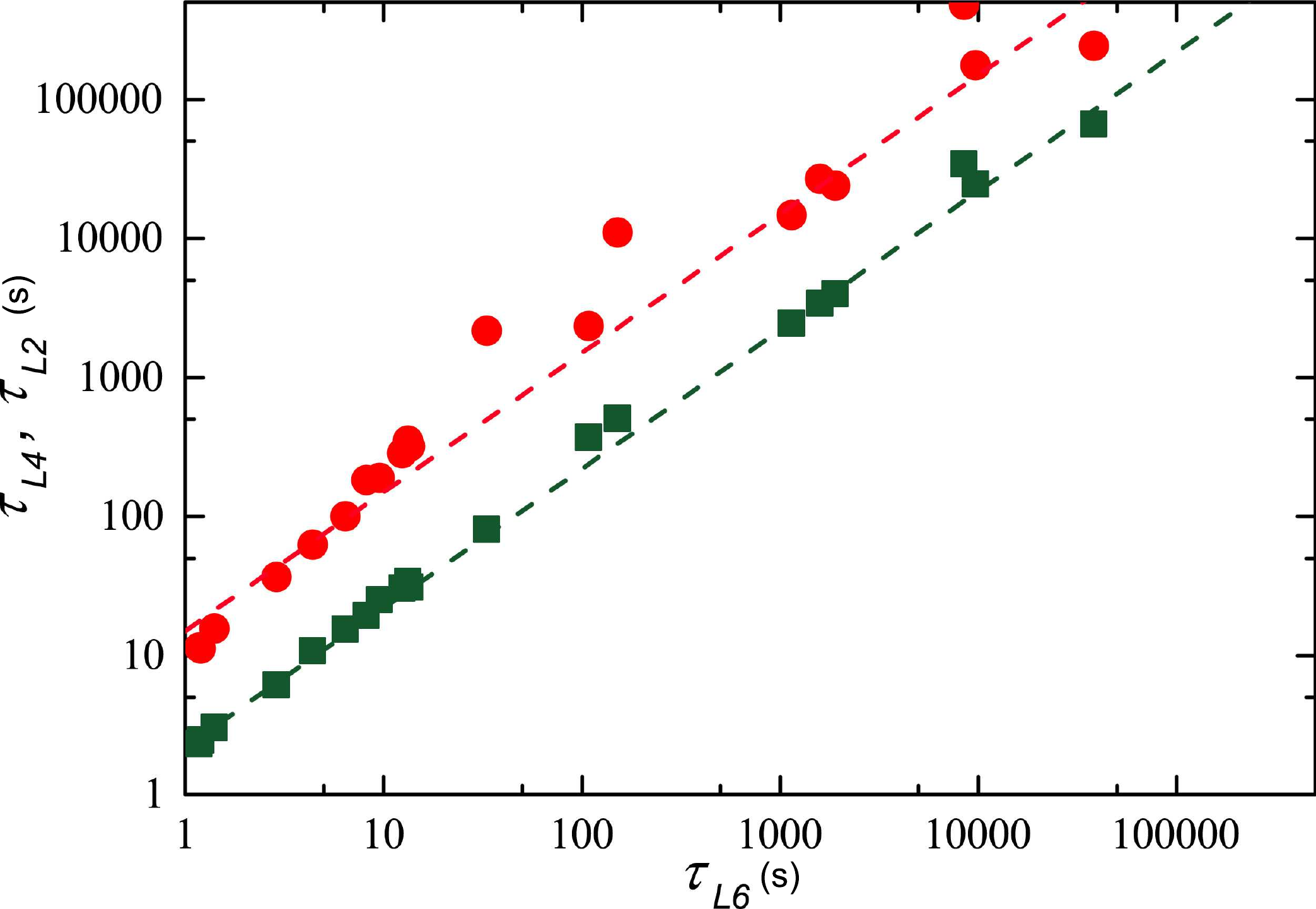}
	\end{center}
	\caption[Relaxation of the body orientational correlation function]
	{Relaxation of the body orientational correlation function $L_n$. Relaxation time of $L_2$, $\tau_{L2}$ (red disks) and the relaxation time of $L_4$, $\tau_{L4}$ (green squares), as a function of the relaxation time of $L_6$, $\tau_{L6}$. The red dashed line indicates $\tau_{L2} = 15\tau_{L6}$. The green dashed line indicates $\tau_{L4} = 2.2\tau_{L6}$. 
	}
	\label{Figure12}
\end{figure}

The relaxation of the orientational degree of freedom of ellipsoids can be characterized by the body orientational correlation function, $L_n$ (Eq.~\ref{body-orientation}). Zheng {\it et al.} has measured $L_n$ of ellipsoids in 2D.\cite{Zheng11,Zheng14} They showed that although $L_n$ decays faster for larger $n$ as expected, different choices of $n$ yield the same glass transition point. We verify their finding further by examining the relaxation of $L_n$ with $n = 2$, $4$ and $6$. The relaxation times of $L_2$, $L_4$ and $L_6$ are linearly proportional to each other (Fig.~\ref{Figure12}). Hence, these different choices of $n$ should not affect the conclusion of our paper. We choose the large $n = 6$ in our study so that the body orientational relaxation time can be measured at slightly higher area fractions. Note that $L_2$ decays 15 times slower than that of $L_6$ at a given $\phi$, whereas $L_4$ decays 2.2 times slower than that of $L_6$. At high area fractions, the data of $L_2$ becomes significantly more noisy due to insufficient statistics at long relaxation times (Fig.~\ref{Figure12}). 

\subsection*{C.2 Two-step glass transition}  

Zheng and co-workers identified the translational and orientational glass transition by extrapolating the corresponding translational and orientational relaxation times to infinity to extract the ideal glass transition points.\cite{Zheng11,Zheng14} Following their procedure, we fit the relaxation times $\tau_F$ and $\tau_L$, which are extracted from the translational correlation $F_s$ and the body orientational correlation $L_6$ respectively, as a function of $\phi$ using power-law relations:
\begin{equation}
\tau(\phi) \propto \left( \phi_c-\phi \right)^{-\gamma} 
\end{equation}
(see Eq. (1) of Ref. 12). Here, $\phi_c$ is the ideal glass transition point, where $\tau$ diverges.  

For the low aspect ratio ($P = 1.9$) ellipsoids, we find the difference between $\phi_c$ from $\tau_F(\phi)$ and $\phi_c$ from $\tau_L(\phi)$ is small: $\phi_c=0.832$ from $\tau_F(\phi)$, whereas $\phi_c=0.826$ from $\tau_L(\phi)$. This is consistent with the study of Zheng and co-workers, where they found the difference in $\tau_c$ goes to zero when $P \le 2.5$ in experiments and when $P \le 1.7$ in simulations. For the high aspect ratio ($P = 6.7$) ellipsoids, we find that $\phi_c=0.85$ from the translational relaxation $\tau_F(\phi)$ and $\phi_c=0.77$ from orientational relaxation $\tau_L(\phi)$. The finite difference between $\phi_c$ from the translational and orientational relaxations, $\Delta \phi_c=0.08$, directly confirms the existence of the two-step glass transition in our experiments. In comparison, Zheng {\it et al.} found $\phi_c=0.78$ from $\tau_F(\phi)$ and $\phi_c=0.70$ from $\tau_L(\phi)$ with ellipsoids of a similar aspect ratio. Thus, $\Delta \phi_c=0.08$ in their study. Although the absolute values of our $\phi_c$ are consistently larger, the differences, $\Delta \phi_c$, are quantitatively the same in the two studies.



\balance



\end{document}